\documentclass[epj,nopacs]{svjour}
\usepackage[intlimits]{amsmath}
\usepackage{graphicx}
\def\makeheadbox{{%
\hbox to0pt{\vbox{\baselineskip=10dd\hrule\hbox
to\hsize{\vrule\kern3pt\vbox{\kern3pt
\hbox{\bfseries Preprint hep-ph/0603133}
\hbox{TTP06-05, PSI-PR-06-06, SFB/CPP-06-07}
\kern3pt}\hfil\kern3pt\vrule}\hrule}%
\hss}}}

\graphicspath{{}{fig/}}

\allowdisplaybreaks

\DeclareMathOperator{\Tr}{Tr}
\DeclareMathOperator{\Rep}{Re}

\let\eps=\varepsilon
\newcommand{\Ac}{\mathcal{A}}
\newcommand{\Fc}{\mathcal{F}}
\newcommand{\Oc}{\mathcal{O}}
\newcommand{\Br}{\mathrm{B}}
\newcommand{\vr}{\mathrm{v}}
\newcommand{\MSbar}{\ensuremath{\overline{\text{MS}}}}
\newcommand{\bare}{\mathrm{bare}}
\newcommand{\lqm}{\mathcal{L}}

\newcommand{\valignbox}[2][]{\begin{tabular}[#1]{@{}c@{}} #2 \end{tabular}}
\newcommand{\vcentergraphics}[2][]{%
  \valignbox{\raisebox{-2ex}{\includegraphics[#1]{#2}}}}
\newcommand{\dslash}{\makebox[0pt][l]{\hspace*{0.325em}\makebox[0pt]{/}}}
\newcommand{\Li}[1]{\mathrm{Li}_{#1}}
\newcommand{\Cl}[1]{\mathrm{Cl}_{#1}}
\newcommand{\dd}{\mathrm{d}}
\newcommand{\loopint}[2]{\int\!\frac{\dd^{#1}#2}{(2\pi)^{#1}}}
\newcommand{\loopintf}[2]{\int\!\frac{\dd^{#1}#2}{i\pi^{#1/2}}}
\newcommand{\MBint}[1]{\int_{-i\infty}^{i\infty}\!\frac{\dd #1}{2\pi i}}

\newcommand{\LA}{\mathrm{LA}}
\newcommand{\NP}{\mathrm{NP}}
\newcommand{\BE}{\mathrm{BE}}
\newcommand{\fc}{\mathrm{fc}}
\newcommand{\BECA}{{\ensuremath{\mathrm{BE}C_{\!A}}}}
\newcommand{\Wc}{\mathrm{Wc}}
\newcommand{\WWcc}{{\ensuremath{W\!W\!cc}}}
\newcommand{\WH}{{\ensuremath{W\!H}}}
\newcommand{\Hphi}{{\ensuremath{H\!\phi}}}
\newcommand{\phiphi}{{\ensuremath{\phi\phi}}}
\newcommand{\Tone}{\mathrm{T1}}
\newcommand{\Ttwo}{\mathrm{T2}}
\newcommand{\ToneCA}{{\ensuremath{\mathrm{T1}C_{\!A}}}}
\newcommand{\Higgs}{\mathrm{Higgs}}
\newcommand{\NA}{{C_F C_A+\Higgs}}

\begin{document}

\title{The two-loop vector form factor in the Sudakov limit}

\author{Bernd Jantzen\inst{1,2}\fnmsep
  \thanks{Bernd Feucht in publications before 2005}
  \and Vladimir A. Smirnov\inst{3,4}}
\institute{Paul Scherrer Institut, 5232 Villigen PSI, Switzerland,
    \email{physics@bernd-jantzen.de}
  \and Institut f\"ur Theoretische Teilchenphysik,
    Universit\"at Karlsruhe, 76128 Karlsruhe, Germany
  \and Nuclear Physics Institute of Moscow State University,
    119992 Moscow, Russia,
    \email{smirnov@theory.sinp.msu.ru}
  \and II. Institut f\"ur Theoretische Physik,
    Universit\"at Hamburg, 22761 Hamburg, Germany}

\date{\mbox{}\\ \mbox{}}

\abstract{
  Recently two-loop electroweak corrections to the neutral current
  four-fermion processes at high energies have been presented.
  The basic ingredient of this calculation is the evaluation of the
  two-loop corrections to the Abelian vector form factor in a
  spontaneously broken $SU(2)$ gauge model.
  Whereas the final result and the derivation of the four-fermion
  cross sections from evolution equations have been published earlier,
  the calculation of the form factor from the two-loop Feynman
  diagrams is presented for the first time in this paper.
  We describe in detail the individual contributions to the form factor
  and their calculation with the help of the expansion by regions
  method and Mellin--Barnes representations.
}

\maketitle

\section{Introduction}

Electroweak higher order corrections in the high energy Sudakov regime
\cite{Sudakov:1954sw,Jackiw:1968} have recently attracted a new wave
of interest
\cite{Kuroda:1990wn,Degrassi:1992ue,Beccaria:1998qe,Ciafaloni:1998xg,%
  Kuhn:1999de,Fadin:1999bq,Kuhn:1999nn,Kuhn:2000hx,%
  Kuhn:2001hz,Kuhn:2001hzE,%
  Beccaria:1999xd,Beccaria:1999fk,Beccaria:2000jz,Beccaria:2001yf,%
  Denner:2000jv,Denner:2001gw,Melles:2000ia,%
  Hori:2000tm,Beenakker:2001kf,%
  Feucht:2003yx,%
  Denner:2003wi,Beccaria:2003ct,Beccaria:2003yn,Pozzorini:2004rm,%
  Feucht:2004rp,%
  Kuhn:2004em,Kuhn:2005az,Kuhn:2005gv,%
  Accomando:2004de,Denner:2004iz,%
  Jantzen:2005xi,Jantzen:2005az}.
At the upcoming colliders, the LHC and an International
Linear Collider, for the first time the characteristic energies
$\sqrt s$ of the partonic processes will be far larger than the masses
of the $W$- and $Z$-bosons, $M_{W,Z}$.
In view of the expected experimental accuracy, one has to take into
account radiative corrections at the two-loop level which are enhanced
by up to four powers of the large electroweak logarithm
$\ln(s/M_{W,Z}^2)$.
These are present in virtual corrections to exclusive reactions like
electron--positron or quark--antiquark annihilation into a pair of
fermions or gauge bosons.

For the high energy behaviour of the neutral current four-fermion
processes the analysis of the leading logarithms (LL)
in \cite{Fadin:1999bq} was extended to the next-to-leading (NLL)
and next-to-next-to-leading logarithmic (NNLL) level in 
\cite{Kuhn:1999nn,Kuhn:2000hx,Kuhn:2001hz,Kuhn:2001hzE}.
With the help of evolution equations which describe the dependence of
the amplitude on the energy, the logarithmic corrections were
resummed to all orders in perturbation theory in NNLL accuracy.
Neglecting the fermion masses and the mass difference between the $W$-
and $Z$-boson, the logarithmically enhanced part of the two-loop
corrections to the total cross section and to various asymmetries was
obtained including the $\ln^n(s/M_{W,Z}^2)$ terms with $n=4,3,2$.
The results up to NLL accuracy have been confirmed by explicit
one-loop
\cite{Beccaria:1998qe,Beccaria:2000jz,Denner:2000jv,Denner:2001gw}
and two-loop
\cite{Hori:2000tm,Beenakker:2001kf,Denner:2003wi,Pozzorini:2004rm}
calculations.

For energies in the TeV region the subleading logarithmic
contributions are comparable in size to the leading terms due to their
large numerical coefficients.
Thus the calculation of the remaining two-loop linear logarithms is
necessary to control the convergence of the logarithmic expansion.
These corrections represent the next-to-next-to-next-to-leading
logarithmic (N$^3$LL) contributions.
In contrast to the higher powers of the electroweak logarithm, they
are sensitive to the details of the gauge boson mass generation, in
particular they depend on the Higgs boson mass $M_H$.
Thanks to the evolution equations, the NNLL calculation involves only
massless Feynman diagrams at the two-loop level. But the linear
two-loop logarithm requires the evaluation of vertex corrections with
the true masses of the gauge bosons and the Higgs boson.

In \cite{Feucht:2003yx,Feucht:2004rp,Jantzen:2005xi,Jantzen:2005az}
the previous analysis is extended to N$^3$LL accuracy. The application
of the evolution equation approach to the linear two-loop logarithm
and the necessary ingredients are described in detail
in~\cite{Jantzen:2005az}.
{}From the viewpoint of loop calculations, the most complicated
contributions are the massive two-loop corrections to the Abelian
vector form factor which will be defined in section~\ref{sec:formfactor}.
In \cite{Jantzen:2005xi,Jantzen:2005az} the form factor results are
used together with the evolution equations to obtain the N$^3$LL
two-loop corrections to the four-fermion scattering amplitude in a
spontaneously broken $SU(2)$ gauge model.
The additional infrared-diver\-gent electromagnetic contributions are
separated according to the prescription developed in
\cite{Feucht:2004rp}.
Finally the effect of the mass difference between the two heavy
electroweak gauge bosons $W$ and $Z$ is taken into account by an
expansion around the equal mase case, whereas a value of the Higgs
mass identical to $M_W$ is sufficient for the desired accuracy.
In this way electroweak corrections to the total cross section,
forward--backward asymmetries and left--right asymmetries of the neutral
current four-fermion processes are obtained including all
large two-loop logarithms and leaving an estimated theoretical
uncertainty of a few per mil to one percent for the production of
light fermions~\cite{Jantzen:2005xi,Jantzen:2005az}.

This calculation and the discussion of the results are not
repeated here. The following sections are instead dedicated to details
of the loop calculations needed for the form factor mentioned above.
The paper is organized as follows.
In section~\ref{sec:formfactor} the Abelian vector form factor in the
spontaneously broken $SU(2)$ gauge model is defined.
Then section~\ref{sec:vertex} describes the evaluation of the two-loop
vertex corrections to the form factor. Contributions from the
renormalization of the fields, the coupling constant and the gauge
boson mass are added in section~\ref{sec:ren}.
Finally we discuss the result for the form factor in
section~\ref{sec:result} and conclude with a summary in
section~\ref{sec:summary}.
The appendices list the Feynman rules for the $SU(2)$ model
(appendix~\ref{sec:feynman}) and explain two important methods used in
our calculation: the expansion by regions (appendix~\ref{sec:regions})
and the Mellin--Barnes representation (appendix~\ref{sec:MB}).
At last appendix~\ref{sec:massgap} lists the contributions in a
theory with a mass gap which are necessary for the separation of the
electromagnetic corrections.

\section{The Abelian vector form factor}
\label{sec:formfactor}

The Abelian vector form factor~$F$ determines the fermion scattering
in an external Abelian field. It is the factor which multiplies the
Born term $\Fc_\Br^\mu = \bar\psi(p_1) \gamma^\mu \psi(p_2)$
in the corrections to the Abelian vector current
$\Fc^\mu = F(Q^2) \Fc_\Br^\mu$, where $p_1$ denotes
the outgoing and $p_2$ the incoming fer\-mion momentum and
$Q^2 = -(p_1-p_2)^2$.
At high energies, we consider the Sudakov limit
\cite{Sudakov:1954sw,Jackiw:1968} $Q^2 \to \infty$, so $Q^2 \gg M^2$
for every gauge boson or Higgs mass~$M$, we neglect fermion masses,
$p_1^2 = p_2^2 = 0$, and we omit terms which are power-suppressed by
at least one factor $M^2/Q^2$.

The four-fermion amplitude~$\Ac$ describes the neutral current
scattering $f \bar f \to f' \bar f'$ of a fermion--antifermion pair
into a different fermion--antifermion pair.
At high energies and fixed angles, where all kinematical invariants
are of the same order and far larger than the gauge boson mass,
$s \sim |t| \sim |u| \gg M^2$, it can be decomposed into the form factor
squared and a reduced amplitude~$\tilde\Ac$,
\begin{equation}
\label{eq:Adecomp}
  \Ac = \frac{i g^2}{s} \, F^2 \, \tilde\Ac
  \,,
\end{equation}
where $g$ is the weak $SU(2)$ coupling.
The collinear divergences appearing in the limit of vanishing gauge
boson mass, $M \to 0$, and the corresponding collinear logarithms are
known to factorize. They are responsible, in particular, for the
double-logarithmic contribution and depend only on the properties of
the external on-shell particles, but not on the specific process
\cite{Cornwall:1975aq,Cornwall:1975ty,Frenkel:1976bj,Amati:1978by,%
  Mueller:1979ih,Collins:1980ih,Collins:1989bt,Sen:1981sd,Sen:1982bt}.
Thus, for each fermion--antifermion pair of the four-fermion amplitude
the collinear logarithms are the same as for the form factor~$F$
discussed above.
The reduced amplitude~$\tilde\Ac$ in~(\ref{eq:Adecomp}) therefore
contains only soft logarithms (corresponding to soft divergences in the
limit $M \to 0$) and renormalization group logarithms. It can be
determined with the help of an evolution equation
\cite{Sen:1982bt,Sterman:1986aj,Botts:1989kf}.

The decomposition of the four-fermion amplitude~$\Ac$ and the
calculation of the reduced amplitude~$\tilde\Ac$ are described in
detail in \cite{Kuhn:1999nn,Kuhn:2000hx,Kuhn:2001hz,Kuhn:2001hzE}.
The NNLL approximation of the two-loop contribution to the form
factor~$F$ can be obtained with the help of another evolution
equation~\cite{Kuhn:2001hz,Kuhn:2001hzE}, whereas the N$^3$LL result
including all large logarithms requires the two-loop calculations with
massive gauge bosons presented in the following sections.

The form factor is calculated as a real function of the variable
$Q^2 > 0$, i.e. in the Euclidean region. For its application to the
four-fermion amplitude described above, the analytic continuation to
the Minkowskian region $s > 0$ according to $Q^2 = -(s + i0)$,
where $i0$ denotes an infinitesimal positive imaginary part,
leads to the substitution $\ln(Q^2/M^2) = \ln(s/M^2) - i\pi$.

The calculation is performed in a spontaneously broken $SU(2)$
gauge model. Reference \cite{Jantzen:2005az} discusses in detail all
effects resulting from the difference of this model with respect to
the standard model of particle physics, where the isospin $SU(2)$
group for left-handed fermions is mixed with the hypercharge $U(1)$
group through the mass eigenstates of the $Z$-boson and the photon.

In contrast to the standard model particles $W^\pm$ and $Z$, we work
with the neutral $SU(2)$ gauge bosons $W^a$, $a=1,2,3$, which all have
the same mass $M=M_W$.
The generators of an $SU(N)$ gauge group in the fundamental
representation are labelled $t^a$. Their Lie algebra
involves the structure constants $f^{abc}$.
The Casimir operators of the fundamental and the adjoint
representation are $C_F = (N^2-1)/(2N)$ and $C_A = N$, respectively.
In addition $T_F = 1/2$ is needed.
In the special case of an $SU(2)$ group the generators $t^a$
correspond to half the Pauli matrices, and $f^{abc} = \eps^{abc}$,
$C_F = 3/4$, $C_A = 2$.
We prefer to use the general symbols $t^a$, $f^{abc}$, $C_F$, $C_A$
and $T_F$ instead of their specific $SU(2)$ values in our
calculations. This has the advantage that we can easily convert the
results to the case of the hypercharge $U(1)$ gauge group.

The Feynman rules of the vertices needed for our calculation are
listed in appendix~\ref{sec:feynman}. We use the Feynman--'t~Hooft
gauge, where the masses of Goldstone bosons and ghost fields are equal
to the gauge boson mass $M$ and the gauge boson propagators have the
form $-i g^{\mu\nu}/(k^2-M^2)$.
We work with the Lagrangian as a function of the unrenormalized
quantities, so instead of calculating diagrams with counter terms, we have
to replace the bare mass and coupling constant in the one-loop result
by the corresponding renormalized quantities as described in
section~\ref{sec:ren}.

\section{Vertex corrections}
\label{sec:vertex}

In this section the two-loop vertex corrections to the Abelian vector
form factor are presented.
The Feynman diagrams contributing to the vertex corrections are
depicted in figures \ref{fig:feynnf}, \ref{fig:feynAb} and
\ref{fig:feynNA}.%
{\newcommand{\fscale}{0.8}%
\newcommand{\fhspace}{\hspace{1em}}\newcommand{\fvspace}{\\[3ex]}%
\begin{figure}
  \centering
  \includegraphics[scale=\fscale]{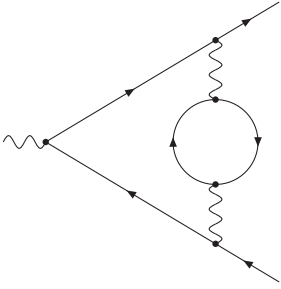}
  \caption{Fermionic vertex correction}
  \label{fig:feynnf}
\end{figure}%
\begin{figure}
  \centering
  \valignbox[b]{\includegraphics[scale=\fscale]{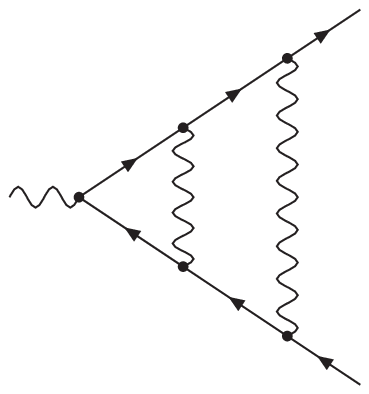}\\(a)}
  \fhspace
  \valignbox[b]{\includegraphics[scale=\fscale]{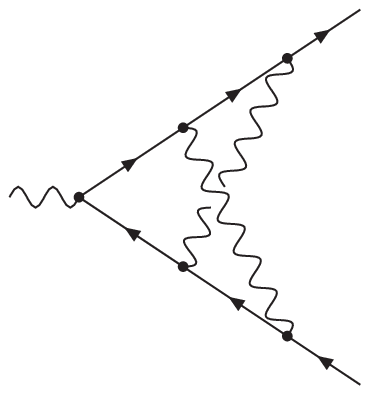}\\(b)}
  \fvspace
  \valignbox[b]{\includegraphics[scale=\fscale]{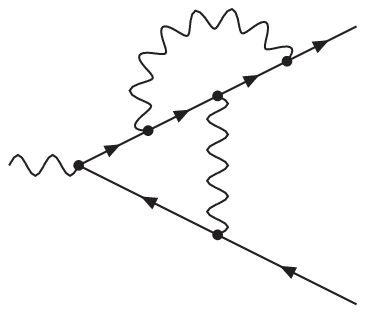}\\(c)}
  \fhspace
  \valignbox[b]{\includegraphics[scale=\fscale]{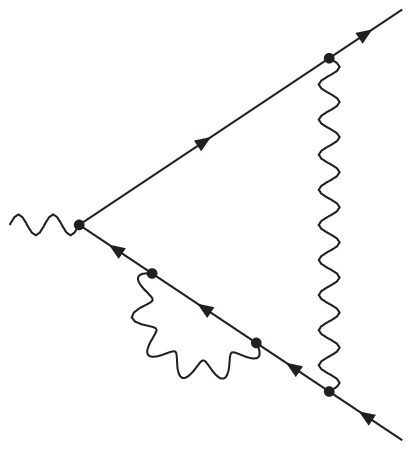}\\(d)}
  \caption{Abelian vertex corrections}
  \label{fig:feynAb}
\end{figure}%
\begin{figure}
  \centering
  \valignbox[b]{\includegraphics[scale=\fscale]{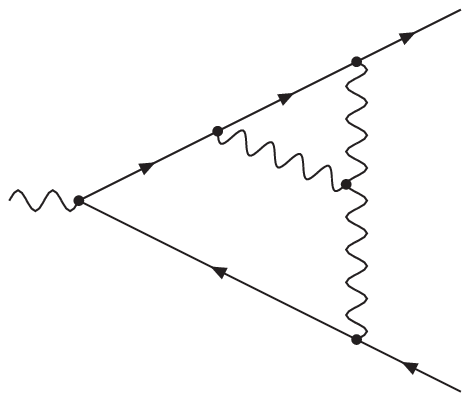}\\(a)}
  \fhspace
  \valignbox[b]{\includegraphics[scale=\fscale]{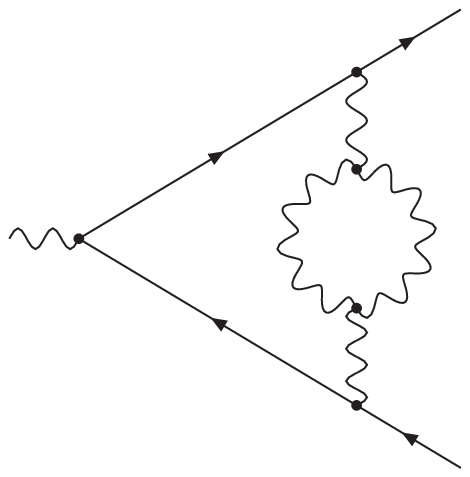}\\(b)}
  \fvspace
  \valignbox[b]{\includegraphics[scale=\fscale]{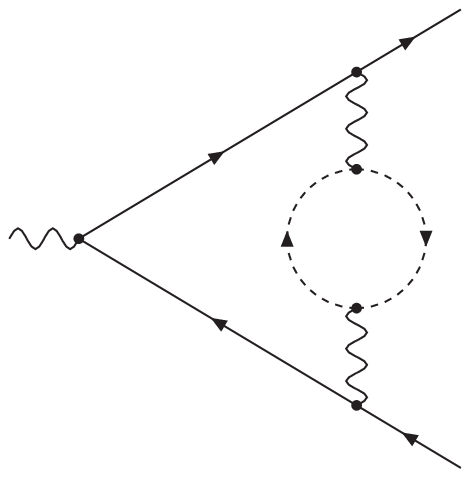}\\(c)}
  \fhspace
  \valignbox[b]{\includegraphics[scale=\fscale]{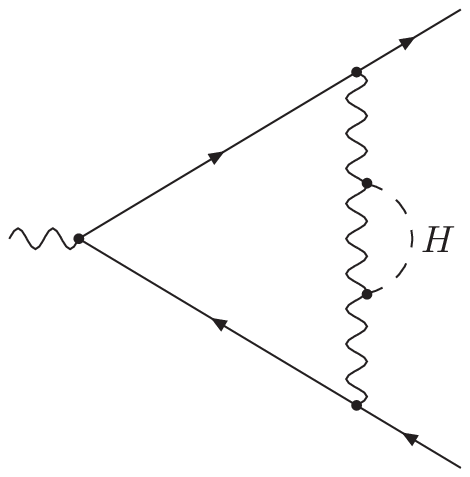}\\(d)}
  \fvspace
  \valignbox[b]{\includegraphics[scale=\fscale]{%
    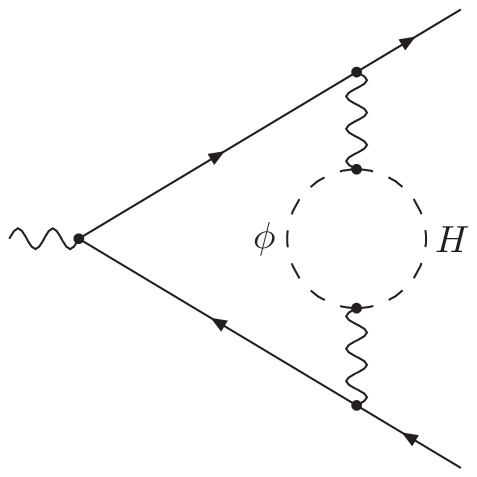}\\(e)}
  \fhspace
  \valignbox[b]{\includegraphics[scale=\fscale]{%
    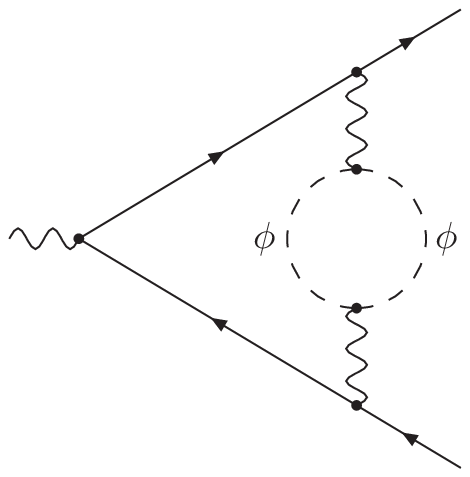}\\(f)}
  \caption{Non-Abelian vertex corrections}
  \label{fig:feynNA}
\end{figure}%
}
Solid lines with arrows denote fermions, wavy lines denote gauge
bosons, short-dashed lines with arrows denote ghost fields
and long-dashed lines stand for the Higgs boson~($H$) or Goldstone
bosons~($\phi$), depending on the labels.
The graphs in figures \ref{fig:feynAb}c), \ref{fig:feynAb}d) and
\ref{fig:feynNA}a) also appear in a horizontally mirrored version, so
their contributions have to be counted twice.

The three figures group the Feynman diagrams in subsets which are
separately gauge-invariant when adding the corresponding
renormalization contributions from section~\ref{sec:ren}.
The fermionic contribution of the graph in figure~\ref{fig:feynnf} is
proportional to $n_f$, the number of fermions running in the closed
fermion loop.
Figure~\ref{fig:feynAb} represents the Abelian graphs (in
addition to figure~\ref{fig:feynnf}) which are present also in an
unbroken $U(1)$ theory like QED.
Finally figure~\ref{fig:feynNA} shows the non-Abelian graphs,
which include the contributions from the Higgs mechanism.
The Abelian contribution only counts the part of the graphs
\ref{fig:feynAb}b) and \ref{fig:feynAb}c) which is proportional to
$C_F^2$. The other part of these two graphs, which is proportional to
$C_F C_A$, belongs to the non-Abelian contribution.

The fermionic contribution has been calculated exactly
in~\cite{Feucht:2003yx}, i.e. for all $Q^2$, not only $Q^2 \gg M^2$,
showing the good agreement of the Sudakov limit with the exact
contribution for energies larger than 300~GeV.
The high-energy asymptotic limit of this result is quoted in
section~\ref{sec:vertexnf}.
Note that we state in this paper the individual vertex correction,
self-energy correction and renormalization terms, whereas in
\cite{Feucht:2003yx} only the total fermionic contribution to the form
factor is given.

The Abelian graphs have been evaluated in N$^4$LL approximation,
i.e. including all large logarithms \emph{and} the non-logarithmic
constant. These calculations are presented in sections
\ref{sec:vertexLA} to \ref{sec:vertexfc}.
The non-Abelian graphs, especially figure~\ref{fig:feynNA}a), are more
complicated to evaluate, as they have three massive propagators each
(compared to two for the Abelian graphs). We have only evaluated them
in N$^3$LL accuracy as the non-logarithmic constant is not needed for
the insertion of the form factor result into the four-fermion
amplitude. The corresponding calculations can be found in sections
\ref{sec:vertexBECA} and \ref{sec:vertexWc}.

\subsection{Reduction to scalar integrals}
\label{sec:redscalar}

{}From each Feynman vertex diagram in the figures
\ref{fig:feynnf}--\ref{fig:feynNA}, by applying the Feynman rules in
appendix~\ref{sec:feynman}, we get a vertex amplitude of the following form:
\begin{equation}
  \Fc_\vr^\mu = \bar\psi(p_1) \, \Gamma^\mu \, \psi(p_2) \,,
\end{equation}
where $\psi(p_{1,2})$ are doublets of Dirac spinors in the $SU(2)$
isospin space corresponding to the incoming and outgoing fermion, and
$\Gamma^\mu$ is a quadratic matrix both in the spinor space and in the
isospin space (for each Lorentz index~$\mu$).

For vanishing fermion masses (in the Sudakov limit) the vertex
amplitudes can be written as $\Fc_\vr^\mu = F_\vr \Fc_\Br^\mu$, where
$\Fc_\Br^\mu$ is the Born amplitude and $F_\vr$ is a contribution to
the form factor.
The scalar quantity~$F_\vr$ can be extracted from the vertex amplitude
by projection (see e.g. \cite{Bernreuther:2004ih}):
\begin{equation}
  F_\vr = -\frac{\Tr(\gamma_\mu \dslash p_1 \Gamma^\mu \dslash p_2)}{%
             4 N (d-2) \, p_1 \cdot p_2}
  \,,
\end{equation}
where dimensional regularization \cite{'tHooft:1972fi} is used with
$d = 4 - 2\eps$ as the number of space-time dimensions,
and $N=2$ for $SU(2)$.
The trace runs over the spinor and the isospin indices.
By applying this projection, we get a linear combination of scalar
loop integrals.
For convenience, we separate the integration measure as follows:
\begin{equation}
  \mu^{4-d} g^2 \loopint dk =
  i \frac{\alpha}{4\pi} \left(\frac{\mu^2}{M^2}\right)^{\eps} S_\eps
  \left[ e^{\eps\gamma_E} (M^2)^\eps \loopintf dk \right]
\end{equation}
Here $\mu$ is the mass scale of dimensional regularization,
$\alpha = g^2/(4\pi)$ with the weak coupling~$g$,
$\gamma_E$ is Euler's constant
and $S_\eps = (4\pi)^\eps e^{-\eps\gamma_E}$.
Within the \MSbar{} renormalization scheme, $S_\eps$ is absorbed into
$\mu^{2\eps}$ by a redefinition of $\mu$, and as we set $\mu = M$ in
the end, the prefactor in front of the square brackets gets especially
simple.

The reduction of the Feynman amplitudes to scalar integrals has been
performed with the computer algebra program FORM
\cite{Vermaseren:2000nd},
and the evaluation of the scalar integrals, as described in the
following sections, has been done with \textsc{Mathematica}
\cite{Wolfram:Mathematica4.2}.

We have not performed a reduction of the scalar integrals to so-called
master integrals by a method like integration by
parts~\cite{Tkachov:1981wb,Chetyrkin:1981qh}, as the number of scalar
integrals obtained from the Feynman diagrams is not too big and most
of the scalar integrals can easily be evaluated in a semi-automatical
way starting from our expressions for general powers of the
propagators, which are presented in the following sections.

\subsection{Fermionic vertex correction}
\label{sec:vertexnf}

The fermionic vertex correction of figure~\ref{fig:feynnf} has been
evaluated in~\cite{Feucht:2003yx}, where the integration of the inner
fermion loop has been done first, leaving a one-loop integral feasible
by the standard Feynman parametrization technique.
The contribution of the fermionic vertex correction to the form factor
is
\begin{align}
\label{eq:Fvnfres}
  F_{\vr,n_f} &= C_F T_F n_f \left(\frac{\alpha}{4\pi}\right)^2
    \left(\frac{\mu^2}{M^2}\right)^{2\eps} S_\eps^2 \,
    \Biggl\{
  \nonumber \\* & \qquad
      \frac{1}{\eps} \left[ \frac{4}{3} \lqm^2 - \frac{20}{3} \lqm
        + \frac{8}{9}\pi^2 + \frac{29}{3} \right]
    - \frac{8}{9} \lqm^3
    + \frac{56}{9} \lqm^2
  \nonumber \\ & \qquad
    + \left(\frac{4}{9}\pi^2 - \frac{238}{9}\right) \lqm
    - \frac{8}{3}\zeta_3 - \frac{38}{27}\pi^2 + \frac{749}{18}
    \Biggr\}
  \nonumber \\* & \quad
  + \Oc(\eps) + \Oc\!\left(\frac{M^2}{Q^2}\right) ,
\end{align}
where $\lqm = \ln(Q^2/M^2)$, and $\zeta_3 \approx 1.202057$ is a value
of Riemann's zeta function.

\subsection{Planar vertex correction}
\label{sec:vertexLA}

\begin{figure}
  \centering
  \includegraphics{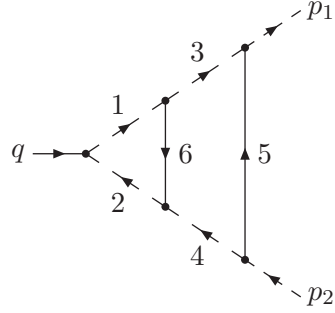}
  \caption{Scalar graph for planar vertex correction}
  \label{fig:scalarLA}
\end{figure}

The reduction (see section~\ref{sec:redscalar}) of
the planar Feynman graph in figure~\ref{fig:feynAb}a) leads to scalar
integrals corresponding to the graph in figure~\ref{fig:scalarLA}.%
The numbers enumerate the inner propagators and correspond to the
indices~$i$ of the propagator powers~$n_i$ and of the inner
momenta~$k_i$, the directions of which are indicated by the
arrows. Solid lines stand for massive propagators, dashed lines for
massless ones.

Apart from propagators in the denominator, one scalar product remains
in the numerator which cannot be expressed linearly in terms of the
denominator. We have chosen this irreducible scalar product to be
$2k_5 \cdot (k_5-k_6)$.
The set of scalar integrals is then covered by the following function
($k=k_5$, $\ell=k_5-k_6$):
\begin{align}
\label{eq:LAscalar}
  \lefteqn{F_\LA(n_1,\ldots,n_7) =
    e^{2\eps\gamma_E} \, (M^2)^{2\eps} \, (Q^2)^{n-n_7-4}}
  \nonumber \\* & \times
  \loopintf dk \loopintf d\ell \,
  \frac{(2k \cdot \ell)^{n_7}}{(\ell^2 - 2p_1\cdot\ell)^{n_1} \,
    (\ell^2 - 2p_2\cdot\ell)^{n_2}}
  \nonumber \\* & \times
  \frac{1}{(k^2 - 2p_1\cdot k)^{n_3} \, (k^2 - 2p_2\cdot k)^{n_4} \,
    (k^2 - M^2)^{n_5}}
  \nonumber \\* &
  \times \frac{1}{((k-\ell)^2 - M^2)^{n_6}}
  \,,
\end{align}
with $n = n_{123456}$, where we use $n_{ij\cdots} = n_i + n_j + \ldots$
as a shorthand notation.
The scalar integrals have been defined in such a way that they do not
carry a mass dimension.

The evaluation of the scalar integrals has been performed with the
expansion by regions (see appendix~\ref{sec:regions}).
The following regions contribute to the planar vertex correction
\cite{Smirnov:2002pj}:
\begin{alignat*}{6}
  \text{(h-h):} & \quad && k \sim Q, \quad && \ell \sim Q \\
  \text{(1c-h):} &&& k \;\|\; p_1, && \ell \sim Q \\
  \text{(2c-h):} &&& k \;\|\; p_2, && \ell \sim Q \\
  \text{(1c-1c):} &&& k \;\|\; p_1, && \ell \;\|\; p_1 \\
  \text{(2c-2c):} &&& k \;\|\; p_2, && \ell \;\|\; p_2 \\
  \text{(h-s'):} &&& k \sim Q, && k_6 = k-\ell \sim M
\end{alignat*}
By $k \sim Q$ we mean that each component of the vector~$k$ is of the
order of $Q$. And $k \;\|\; p_i$ indicates a region where the
momentum~$k$ is collinear to the external momentum~$p_i$:
\begin{align}
\label{eq:def1c}
  k \;\|\; p_1 &\iff k_+ \sim \frac{M^2}{Q} \,,\quad
    k_- \sim Q \,,\quad k_\bot \sim M
    \,, \\
\label{eq:def2c}
  k \;\|\; p_2 &\iff k_+ \sim Q \,,\quad
    k_- \sim \frac{M^2}{Q} \,,\quad k_\bot \sim M
  \,,
\end{align}
where $k_\pm = (2p_{1,2} \cdot k)/Q$ denotes the components of
$k$ in the direction of $p_2$ and $p_1$ respectively, and
the vector $k_\bot = k - (k_-/Q) p_1 - (k_+/Q) p_2$ is made up of
the components of $k$ perpendicular to $p_{1,2}$.

The leading term in the expansion of the (h-h) region corresponds to
the massless integral with $M=0$, which is well known
\cite{Kramer:1986sg,Kramer:1986sgE,Matsuura:1988sm}.
The (h-s') region is of order $(M^2/Q^2)^{2-n_6+\eps}$ and therefore
suppressed by at least one factor $M^2/Q^2$ with respect to the (h-h)
region for all scalar integrals we need, i.e. for $n_i \le 1$,
$i=1,\ldots,6$. So we do not need to consider the (h-s') region.
The (1c-1c) region is of order $(M^2/Q^2)^{4-n_{1356}+n_7}$, the
(2c-2c) region of order $(M^2/Q^2)^{4-n_{2456}+n_7}$. Both are
suppressed if $n_7 > 0$, i.e. if the numerator is present, and are
only evaluated for $n_7 = 0$.
The leading contributions from the (1c-h) and (1c-1c) regions can be
expressed by one- and two-fold Mellin--Barnes representations (see
appendix~\ref{sec:MB}):
\begin{align}
\label{eq:FLA1chgen}
  \lefteqn{F_\LA^{\text{(1c-h)}}(n_1,\ldots,n_7) =
    \left(\frac{M^2}{Q^2}\right)^{2-n_{35}+\eps} \,
    e^{-n i\pi} \,
    e^{2\eps\gamma_E}}
  \nonumber \\* &
  \times \frac{\Gamma(\frac d2-n_3) \Gamma(\frac d2-n_{16}+n_7)
    \Gamma(n_{35}-\frac d2)}{
    \Gamma(n_1) \Gamma(n_2) \Gamma(n_3) \Gamma(n_5) \Gamma(n_6)
    \Gamma(d-n_{126}+n_7)}
  \nonumber \\* & \times
  \MBint z \, \Gamma(-z) \Gamma(\tfrac d2-n_{26}-z)
  \nonumber \\* & \times
  \frac{\Gamma(n_6+z) \Gamma(n_{37}-n_4+z) \Gamma(n_{126}-\frac d2+z)}{
      \Gamma(\frac d2-n_4+n_7+z)}
  \,,
\\
\label{eq:FLA1c1cgen}
  \lefteqn{F_\LA^{\text{(1c-1c)}}(n_1,\ldots,n_6,n_7=0) =
    \left(\frac{M^2}{Q^2}\right)^{4-n_{1356}} \,
    e^{-n i\pi} \, e^{2\eps\gamma_E}}
  \nonumber \\ & \times
  \frac{1}{
    \Gamma(n_1) \Gamma(n_3) \Gamma(n_5) \Gamma(n_6) \Gamma(\frac d2-n_{24})}
  \MBint{z_1} \MBint{z_2}
  \nonumber \\ & \times
  \frac{\Gamma(-z_1) \Gamma(n_{13}-\frac d2-z_1)
    \Gamma(\frac d2-n_1+z_1)}{
    \Gamma(\frac d2-n_4+z_1)}
  \nonumber \\* & \times
  \Gamma(\tfrac d2-n_{24}+z_1) \,
  \frac{\Gamma(-z_2) \Gamma(\frac d2-n_{35}-z_2)}{
    \Gamma(\frac d2-n_5-z_2)}
  \nonumber \\* & \times
  \Gamma(\tfrac d2-n_{45}-z_2) \Gamma(n_{1356}-d+z_2)
  \Gamma(n_5+z_1+z_2)
  \,.
\end{align}
For symmetry reasons we get
$F_\LA^{\text{(2c-h)}}$ from $F_\LA^{\text{(1c-h)}}$
and $F_\LA^{\text{(2c-2c)}}$ from $F_\LA^{\text{(1c-1c)}}$
by exchanging $n_1 \leftrightarrow n_2$ and $n_3 \leftrightarrow n_4$.

The integration contour of the Mellin--Barnes integrals runs from
$-i\infty$ to $+i\infty$ in such a way that poles from gamma
functions of the form $\Gamma(\ldots+z)$ lie on the left hand side of
the contour (``left poles'') and poles from gamma functions of the
form $\Gamma(\ldots-z)$ lie on the right hand side of the contour
(``right poles'').

The Mellin--Barnes integrals in (\ref{eq:FLA1chgen}) and
(\ref{eq:FLA1c1cgen}) are solved by closing the integration contours
either at positive or negative real infinity and summing over the
residues within the contour.
The integrals develop singularities at points in the parameter space
of the $n_i$ where a left pole and a right pole glue together in one
point.
Some of these singularities are cancelled by zeros originating from
gamma functions in the denominator, e.g. in $F_\LA^{\text{(1c-h)}}$
when $n_6 = 0$. Here the result is given by the limit $n_6 \to 0$ to
which only the residue of the integrand at $z=0$ or $z=-n_6$
contributes.

Other singularities in the parameter space are cancelled between
several regions. This is the case for the pole $1/(n_3-n_4)$ which is
cancelled between the (1c-h) and the (2c-h) regions.
Another pole $1/(n_{13}-n_{24})$ is cancelled between the (1c-1c) and
the (2c-2c) regions.
Such singularities, which are regularized analytically with the
parameters~$n_i$ in individual regions, are typical for collinear
regions in the Sudakov limit.
The sum of the contributions from all regions is well-defined
in the framework of dimensional regularization.

In some cases, the first Barnes lemma (see appendix~\ref{sec:MB}) is
used to solve one of the two Mellin--Barnes integrations
in~(\ref{eq:FLA1c1cgen}). In more complicated cases first all residues
which produce singularities are extracted, and the limits of the
analytic regularization and of dimensional regularization ($\eps \to
0$) are performed before summing up the remaining residues. These sums
are then solved by \textsc{Mathematica} or looked up in a summation
table (e.g. in \cite{Smirnov:2004ym}).

By adding together the contributions from all regions we have obtained
the results for all scalar integrals originating from the reduction of
the planar Feynman diagram.
As examples, we show the results for the scalar graph with all
propagators present and various powers of the numerator:
\begin{align}
\label{eq:FLAfullres}
  &\!\! F_\LA(1, 1, 1, 1, 1, 1, 0) =
    \frac{1}{24}\lqm^4 + \frac{\pi^2}{3}\lqm^2 - 6\zeta_3\lqm
    + \frac{31}{180}\pi^4 \,,
\\
  &\!\! F_\LA(1, 1, 1, 1, 1, 1, 1) =
    \frac{\pi^2}{3}\lqm - 10\zeta_3 \,,
\\
  &\!\! F_\LA(1, 1, 1, 1, 1, 1, 2) =
    \frac{1}{2\eps^2} + \frac{1}{\eps}\left(-\lqm + \frac{7}{2}\right)
  \nonumber \\* &
    + \lqm^2
    + \left(\frac{\pi^2}{6} - 8\right)\lqm
    - 11\zeta_3 + \frac{\pi^2}{12} + \frac{37}{2} \,.
\end{align}
Here and for all other results of individual scalar integrals, we omit
the specification ``${}+\Oc(\eps) + \Oc(M^2/Q^2)$'' of the  neglected terms.
The result~(\ref{eq:FLAfullres}) has already been calculated
in~\cite{Smirnov:1997gx}.

The complete Feynman diagram in figure~\ref{fig:feynAb}a) involving
contributions from all scalar integrals with different $n_i$ yields
the following planar vertex correction:
\begin{align}
\label{eq:LAres}
  F_{\vr,\LA} &=
    C_F^2
    \left(\frac{\alpha}{4\pi}\right)^2
    \left(\frac{\mu^2}{M^2}\right)^{2\eps} S_\eps^2
    \, \Biggl\{
  \frac{1}{2\eps^2}
  \nonumber \\* & \qquad
  + \frac{1}{\eps} \left[
    - \lqm^2 + 3 \lqm - \frac{2}{3}\pi^2 - \frac{11}{4} \right]
  + \frac{1}{6} \lqm^4
  \nonumber \\ & \qquad
  + \left(\frac{2}{3}\pi^2 - 1\right) \lqm^2
  + \left(-32\zeta_3 - \pi^2 + \frac{11}{2}\right) \lqm
  \nonumber \\ & \qquad
  + \frac{8}{15}\pi^4 + 62\zeta_3 + \frac{13}{12}\pi^2 - \frac{41}{8}
  \Biggr\}
  \nonumber \\* & \quad
  + \Oc(\eps) + \Oc\!\left(\frac{M^2}{Q^2}\right)
  .
\end{align}

\subsection{Non-planar vertex correction}
\label{sec:vertexNP}

\begin{figure}
  \centering
  \includegraphics{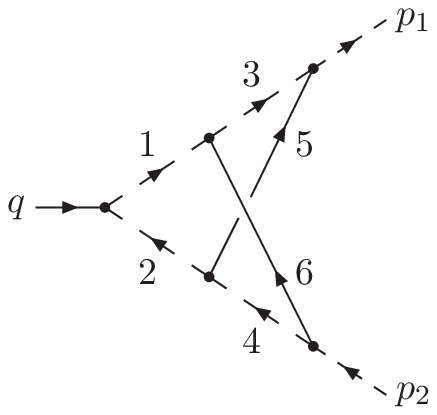}
  \caption{Scalar graph for non-planar vertex correction}
  \label{fig:scalarNP}
\end{figure}

The non-planar Feynman graph in figure~\ref{fig:feynAb}b) involves
the scalar integrals depicted in figure~\ref{fig:scalarNP}.
With the choice $2k_5 \cdot k_6$ for the irreducible scalar product,
the scalar integrals are written as
\begin{align}
\label{eq:NPscalar}
  \lefteqn{F_\NP(n_1,\ldots,n_7) =
    e^{2\eps\gamma_E} \, (M^2)^{2\eps} \, (Q^2)^{n-n_7-4}}
  \nonumber \\* & \times
  \loopintf dk \loopintf d\ell \,
  \frac{(2k \cdot \ell)^{n_7}}{((p_1-k-\ell)^2)^{n_1} \,
    ((p_2-k-\ell)^2)^{n_2}}
  \nonumber \\ & \times
  \frac{1}{(k^2 - 2p_1\cdot k)^{n_3} \,
    (\ell^2 - 2p_2\cdot \ell)^{n_4} \,
    (k^2 - M^2)^{n_5}}
  \nonumber \\* & \times  
  \frac{1}{(\ell^2 - M^2)^{n_6}}
  \,,
\end{align}
with  $k=k_5$ and $\ell=k_6$.%
The following regions contribute to the non-planar vertex
correction \cite{Smirnov:1998vk,Smirnov:2002pj}:
\begin{alignat*}{6}
  \text{(h-h):} & \quad && k \sim Q, \quad && \ell \sim Q \\*
  \text{(1c-h):} &&& k \;\|\; p_1, && \ell \sim Q \\*
  \text{(h-2c):} &&& k \sim Q, && \ell \;\|\; p_2 \\
  \text{(1c-1c):} &&& k \;\|\; p_1, && \ell \;\|\; p_1 \\
  \text{(2c-2c):} &&& k \;\|\; p_2, && \ell \;\|\; p_2 \\
  \text{(1c-2c):} &&& k \;\|\; p_1, && \ell \;\|\; p_2 \\
  \text{(1c-1c'):} &&& k \;\|\; p_1, && k_4 \;\|\; p_1 \\
  \text{(2c'-2c):} &&& k_3 \;\|\; p_2, && \ell \;\|\; p_2 \\
  \text{(us'-us'):} &&& k_3 \sim M^2/Q, \quad && k_4 \sim M^2/Q \\
  \text{(1c-us'):} &&& k \;\|\; p_1, && k_4 \sim M^2/Q \\
  \text{(us'-2c):} &&& k_3 \sim M^2/Q, && \ell \;\|\; p_2
\end{alignat*}
The leading term of the (h-h) region is known from the massless
case \cite{Kramer:1986sg,Kramer:1986sgE,Matsuura:1988sm}.
As for the planar vertex correction in the previous section, the
(1c-1c) and the (2c-2c) regions are of order
$(M^2/Q^2)^{4-n_{1356}+n_7}$ and $(M^2/Q^2)^{4-n_{2456}+n_7}$
respectively. They are suppressed for $n_7>0$ and are therefore
evaluated only for $n_7=0$.
The leading contributions from the regions, apart from (h-h), can be
written as one-fold Mellin--Barnes integrals or simpler expressions:
\begin{align}
\label{eq:FNP1chgen}
  \lefteqn{F_\NP^{\text{(1c-h)}}(n_1,\ldots,n_7) =
    \left(\frac{M^2}{Q^2}\right)^{2-n_{35}+\eps} \,
    e^{-n i\pi} \, e^{2\eps\gamma_E}}
  \nonumber \\* & \times
  \frac{\Gamma(\frac d2-n_{24}) \Gamma(\frac d2-n_{16}+n_7)
    \Gamma(n_{35}-\frac d2)}
    {\Gamma(n_1) \Gamma(n_2) \Gamma(n_3) \Gamma(n_5)
    \, \Gamma(d-n_{1246}+n_7)^2}
    \MBint z
  \nonumber \\ & \times
  \frac{\Gamma(-z) \Gamma(\frac d2-n_{146}-z)
    \Gamma(\frac d2-n_{1246}+n_{37}-z)}
    {\Gamma(\frac d2-n_{16}-z)}
  \nonumber \\* & \times
  \frac{\Gamma(n_1+z) \Gamma(\frac d2-n_3+z) \Gamma(n_{1246}-\frac d2+z)}
    {\Gamma(n_{16}+z)}
  \,,
\\
\label{eq:FNP1c1cgen}
  \lefteqn{F_\NP^{\text{(1c-1c)}}(n_1,\ldots,n_6,n_7=0) =
    \left(\frac{M^2}{Q^2}\right)^{4-n_{1356}}
    e^{-n i\pi} \, e^{2\eps\gamma_E}}
  \nonumber \\* &
  \times \frac{\Gamma(n_{16}-\frac d2) \Gamma(n_{1356}-d)}
    {\Gamma(n_1) \Gamma(n_3) \Gamma(n_5) \Gamma(n_6) \Gamma(\frac d2-n_{24})}
  \MBint z \,
  \Gamma(-z)
  \nonumber \\ & \times
  \Gamma(n_{13}-\tfrac d2-z) \,
  \frac{\Gamma(n_5-n_4+z) \Gamma(\frac d2-n_1+z)}{
    \Gamma(n_{156}-n_4-\frac d2+z)}
  \nonumber \\* & \times
  \frac{\Gamma(\frac d2-n_{24}+z) \Gamma(\frac d2-n_{34}+z)}{
    \Gamma(\frac d2-n_4+z)}
  \,,
\\
\label{eq:FNP1c2cgen}
  \lefteqn{F_\NP^{\text{(1c-2c)}}(n_1,\ldots,n_7) =
    \left(\frac{M^2}{Q^2}\right)^{4-n_{3456}}
    e^{-n i\pi} \, e^{2\eps\gamma_E}}
  \nonumber \\* & \times
  \frac{\Gamma(n_{37}-n_2) \Gamma(n_{47}-n_1)
    \Gamma(\frac d2-n_{13}) \Gamma(\frac d2-n_{24})}
    {\Gamma(n_3) \Gamma(n_4) \Gamma(n_5) \Gamma(n_6)
    \, \Gamma(\frac d2-n_{12}+n_7)^2}
  \nonumber \\* & \times
  \Gamma(n_{35}-\tfrac d2) \Gamma(n_{46}-\tfrac d2)
  \,,
\\
\label{eq:FNP1c1cPgen}
  \lefteqn{F_\NP^{\text{(1c-1c')}}(n_1,\ldots,n_7) =
    \left(\frac{M^2}{Q^2}\right)^{4-n_{2345}}
    e^{-n i\pi} \, e^{2\eps\gamma_E}}
  \nonumber \\* & \times
  \frac{\Gamma(d-n_{234}) \Gamma(n_{24}-\frac d2) \Gamma(n_{2345}-d)}
    {\Gamma(n_2) \Gamma(n_3) \Gamma(n_4) \Gamma(n_5)
    \Gamma(\frac d2-n_{16}+n_7) \Gamma(d-n_{2347})}
  \nonumber \\* & \times
  \MBint z \,
  \Gamma(-z) \Gamma(\tfrac d2-n_{347}-z)
  \Gamma(n_{37}-n_6+z)
  \nonumber \\* & \times
  \frac{\Gamma(\frac d2-n_2+z) \Gamma(\frac d2-n_{16}+n_7+z)}
    {\Gamma(\frac d2-n_6+n_7+z)}
  \,,
\\
\label{eq:FNPusPusPgen}
  \lefteqn{F_\NP^{\text{(us'-us')}}(n_1,\ldots,n_7) =
    \left(\frac{M^2}{Q^2}\right)^{8-n_{1256}-2n_{34}-2\eps} \,
    e^{-n i\pi}}
  \nonumber \\* & \times
  e^{2\eps\gamma_E} \,
  \frac{\Gamma(d-n_{134}) \Gamma(d-n_{234})
    \Gamma(n_{13}-\tfrac d2) \Gamma(n_{24}-\tfrac d2)}
    {\Gamma(n_1) \Gamma(n_2) \Gamma(n_3) \Gamma(n_4) \Gamma(n_5)
    \Gamma(n_6)}
  \nonumber \\* & \times
  \Gamma(n_{2345}-d) \Gamma(n_{1346}-d)
  \,,
\\
\label{eq:FNP1cusPgen}
  \lefteqn{F_\NP^{\text{(1c-us')}}(n_1,\ldots,n_7) =
    \left(\frac{M^2}{Q^2}\right)^{6-n_{2356}-2n_4-\eps} \,
    e^{-n i\pi} \, e^{2\eps\gamma_E}}
  \nonumber \\* & \times
  \frac{\Gamma(\tfrac d2-n_4) \Gamma(\tfrac d2-n_{13})
    \Gamma(d-n_{234})
    \Gamma(n_{24}-\frac d2)}
    {\Gamma(n_2) \Gamma(n_3) \Gamma(n_4) \Gamma(n_5) \Gamma(n_6)
    \Gamma(n_{47}-n_1) \Gamma(\frac d2-n_3)}
  \nonumber \\* & \times
  \Gamma(n_{46}-\tfrac d2) \Gamma(n_{347}-\tfrac d2) \Gamma(n_{2345}-d)
  \,.
\end{align}
Using the symmetry of the non-planar graph under the exchange of the
parameters $n_1 \leftrightarrow n_2$, $n_3 \leftrightarrow n_4$ and
$n_5 \leftrightarrow n_6$, one gets
$F_\NP^{\text{(h-2c)}}$ from $F_\NP^{\text{(1c-h)}}$,
$F_\NP^{\text{(2c-2c)}}$ from $F_\NP^{\text{(1c-1c)}}$,
$F_\NP^{\text{(2c'-2c)}}$ from $F_\NP^{\text{(1c-1c')}}$
and $F_\NP^{\text{(us'-2c)}}$ from $F_\NP^{\text{(1c-us')}}$.
The expression~(\ref{eq:FNPusPusPgen}) for the (us'-us') region is
valid for general~$n_7$ although it does not involve $n_7$ explicitly:
The only dependence on~$n_7$ of this region is cancelled by the
prefactor $(Q^2)^{-n_7}$ in~(\ref{eq:NPscalar}).

We have checked the completeness of our set of regions by writing the
full scalar integral for arbitrary parameters~$n_i$ (except $n_7=0$)
as a four-fold Mellin--Barnes representation.
{}From this expression, we have extracted the residues yielding the
non-suppressed contributions and have found 11 terms with exactly the
same dependence on $M^2/Q^2$ as the 11 regions listed above.

The evaluation of the Mellin--Barnes integrals is done as described in
the previous section.
The structure of singularities needing analytic regularization is more
complicated than in the planar case. Various poles involving
combinations of the parameters~$n_i$ are cancelled between the
collinear regions (1c-1c), (2c-2c), (1c-2c), (1c-1c') and
\mbox{(2c'-2c)}.

The contributions of all regions sum up to the results for the scalar
integrals originating from the reduction of the non-planar Feynman
diagram. Examples of these results are
\begin{align}
\label{eq:FNPfullres}
  &\!\! F_\NP(1, 1, 1, 1, 1, 1, 0) =
    \frac{7}{12}\lqm^4 - \frac{\pi^2}{6}\lqm^2 + 20\zeta_3\lqm
    - \frac{31}{180}\pi^4 \,,
\\
  &\!\! F_\NP(1, 1, 1, 1, 1, 1, 1) =
    \frac{1}{4}\lqm^4 - \frac{\pi^2}{6}\lqm^2 + 14\zeta_3\lqm
    - \frac{\pi^4}{90} \,,
\\
  &\!\! F_\NP(1, 1, 1, 1, 1, 1, 2) =
    \frac{2}{\eps^2} + \frac{1}{\eps}(-4\lqm + 7)
    + \frac{1}{4}\lqm^4 - \lqm^3
  \nonumber \\* &
    + \left(-\frac{\pi^2}{6} + 9\right)\lqm^2
    + (14\zeta_3 - 30)\lqm
    - \frac{\pi^4}{90} - 4\zeta_3+ \frac{\pi^2}{3}
  \nonumber \\* &
    + 38 \,,
\\
  &\!\! F_\NP(1, 1, 1, 1, 1, 1, 3) =
    \frac{7}{2\eps^2} + \frac{1}{\eps}\left(-7\lqm + \frac{111}{8}\right)
  \nonumber \\* &
    + \frac{1}{4}\lqm^4
    - \frac{3}{2}\lqm^3
    + \left(-\frac{\pi^2}{6} + \frac{59}{4}\right)\lqm^2
    + \left(14\zeta_3 - \frac{211}{4}\right)\lqm
  \nonumber \\* &
    - \frac{\pi^4}{90} - 6\zeta_3 + \frac{3}{4}\pi^2 + \frac{571}{8} \,.
\end{align}
The result~(\ref{eq:FNPfullres}) for the scalar graph without numerator
is known from~\cite{Smirnov:1998vk}.
The complete non-planar vertex correction with contributions from all
scalar integrals is as follows:
\begin{align}
\label{eq:NPres}
  F_{\vr,\NP} &=
    \left( C_F^2 - \frac{1}{2} C_F C_A \right)
    \left(\frac{\alpha}{4\pi}\right)^2
    \left(\frac{\mu^2}{M^2}\right)^{2\eps} S_\eps^2
    \, \Biggl\{
  - \frac{2}{\eps}
  \nonumber \\* & \qquad
  + \frac{1}{3} \lqm^4
  - \frac{8}{3} \lqm^3
  + \left(-\frac{2}{3}\pi^2 + 12\right) \lqm^2
  \nonumber \\ & \qquad
  + \left(40\zeta_3 + \frac{2}{3}\pi^2 - 28\right) \lqm
  - \frac{4}{15}\pi^4 - 72\zeta_3 - \pi^2
  \nonumber \\* & \qquad
  + 28
  \Biggr\}
  + \Oc(\eps) + \Oc\!\left(\frac{M^2}{Q^2}\right)
  \,.
\end{align}

\subsection{Vertex correction with Mercedes--Benz graph}
\label{sec:vertexBE}

\begin{figure}
  \centering
  \includegraphics{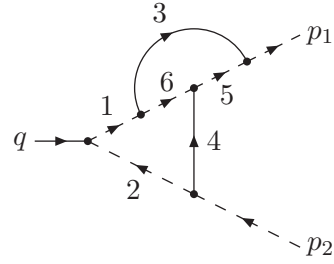}
  \caption{Scalar Mercedes--Benz graph}
  \label{fig:scalarBE}
\end{figure}

Figure~\ref{fig:scalarBE} illustrates the scalar integrals resulting
from the reduction of the Mercedes--Benz graph in
figure~\ref{fig:feynAb}c).%
With our choice of $2p_2 \cdot k_5$ as the irreducible scalar product,
the scalar integrals are defined as
\begin{align}
\label{eq:BEscalar}
  \lefteqn{F_\BE(n_1,\ldots,n_7) =
    e^{2\eps\gamma_E} \, (M^2)^{2\eps} \, (Q^2)^{n-n_7-4}}
  \nonumber \\* & \times
  \loopintf dk \loopintf d\ell \,
  \frac{(2p_2 \cdot k)^{n_7}}
    {(\ell^2-2p_1\cdot\ell)^{n_1} \, (\ell^2-2p_2\cdot\ell)^{n_2}}
  \nonumber \\ & \times
  \frac{1}{(k^2 - 2p_1\cdot k - M^2)^{n_3} \,
    (\ell^2 - M^2)^{n_4} \, (k^2)^{n_5} \, ((k-\ell)^2)^{n_6}}
  \,,
\end{align}
with $k=k_5$ and $\ell=k_4$.
The list of relevant regions for the Mercedes--Benz graph is shown
here:
\begin{alignat*}{6}
    \text{(h-h):} & \quad && k \sim Q, \quad && \ell \sim Q \\
    \text{(1c-h):} &&& k \;\|\; p_1, && \ell \sim Q \\
    \text{(h-2c):} &&& k \sim Q, && \ell \;\|\; p_2 \\
    \text{(us-2c):} &&& k \sim M^2/Q, \quad && \ell \;\|\; p_2 \\
    \text{(1c-1c):} &&& k \;\|\; p_1, && \ell \;\|\; p_1 \\
    \text{(2c-2c):} &&& k \;\|\; p_2, && \ell \;\|\; p_2 \\
    \text{(1c-2c):} &&& k \;\|\; p_1, && \ell \;\|\; p_2
\end{alignat*}
The leading contributions of all regions could be evaluated for
general~$n_7$, but the (us-2c) and the (2c-2c) region are only
non-suppressed for $n_7=0$:
\begin{align}
\label{eq:FBEhhgen}
  \lefteqn{F_\BE^{\text{(h-h)}}(n_1,\ldots,n_7) =
    \left(\frac{M^2}{Q^2}\right)^{2\eps}
    e^{-n i\pi} \, e^{2\eps\gamma_E}
    \sum_{i_1,i_2,i_3\ge0}^{i_{123}\le n_7}}
  \nonumber \\* & \times
  \frac{n_7!}{i_1!\,i_2!\, i_3!\,(n_7-i_{123})!} \,
  \frac{\Gamma(n_1+i_3) \Gamma(n_4+i_2)}
    {\Gamma(n_1) \Gamma(n_2) \Gamma(n_3) \Gamma(n_4)}
  \nonumber \\ & \times
  \frac{\Gamma(\frac d2-n_{35}) \Gamma(d-n_{13456}+i_1) \Gamma(n_{123456}-d)}
    {\Gamma(n_5) \Gamma(n_6)
    \Gamma(d-n_{356}+i_{123}) \Gamma(\frac32d-n_{123456}+n_7)}
  \nonumber \\ & \times
  \MBint z \,
  \Gamma(-z) \Gamma(\tfrac d2-n_{56}+i_{12}-z)
  \Gamma(n_5+z)
  \nonumber \\ & \times
  \Gamma(\tfrac d2-n_{24}+n_7-i_{12}+z) \,
  \frac{\Gamma(n_{3567}-i_{123}-\frac d2+z)}
    {\Gamma(n_{13567}-i_{12}-\frac d2+z)}
  \nonumber \\* & \times
  \frac{\Gamma(d-n_{36}-2n_5+i_{123}-z)}{\Gamma(d-n_{36}-2n_5+i_{12}-z)}
  \,,
\\
\label{eq:FBE1chgen}
  \lefteqn{F_\BE^{\text{(1c-h)}}(n_1,\ldots,n_7) =
    \left(\frac{M^2}{Q^2}\right)^{2-n_{35}+\eps} \,
    e^{-n i\pi} \, e^{2\eps\gamma_E}}
  \nonumber \\* & \times
  \frac{\Gamma(n_{35}-\frac d2) \Gamma(\frac d2-n_{146})}
    {\Gamma(n_1) \Gamma(n_2) \Gamma(n_3) \Gamma(n_6) \Gamma(d-n_{1246})}
  \nonumber \\ & \times
  \MBint z \,
  \Gamma(-z) \Gamma(\tfrac d2-n_{246}-z)
  \nonumber \\ & \times
  \frac{\Gamma(n_6+z) \Gamma(\frac d2-n_5+n_7+z) \Gamma(n_{1246}-\frac d2+z)}
    {\Gamma(\frac d2+n_7+z)}
  \,,
\\
\label{eq:FBEh2cgen}
  \lefteqn{F_\BE^{\text{(h-2c)}}(n_1,\ldots,n_7) =
    \left(\frac{M^2}{Q^2}\right)^{2-n_{24}+\eps} \,
    e^{-n i\pi} \, e^{2\eps\gamma_E}}
  \nonumber \\* & \times
  \frac{\Gamma(\frac d2-n_2) \Gamma(\frac d2-n_{35})
    \Gamma(\frac d2-n_{1356}+n_2)}
    {\Gamma(n_2) \Gamma(n_3) \Gamma(n_4) \Gamma(n_6)}
  \nonumber \\* & \times
  \frac{\Gamma(\frac d2-n_{56}+n_7)
    \Gamma(n_{24}-\frac d2) \Gamma(n_{356}-\frac d2)}
    {\Gamma(d-n_{1356}) \Gamma(d-n_{356}+n_7)}
  \,,
\\
\label{eq:FBEus2cgen}
  \lefteqn{F_\BE^{\text{(us-2c)}}(n_1,\ldots,n_7) =
    \left(\frac{M^2}{Q^2}\right)^{6-n_{2346}-2n_5+n_7-\eps}
    e^{-n i\pi}}
  \nonumber \\* & \times
  e^{2\eps\gamma_E} \,
  \frac{\Gamma(\frac d2-n_5+n_7) \Gamma(d-n_{256}+n_7)
    \Gamma(n_{35}-\tfrac d2)}
    {\Gamma(n_2) \Gamma(n_3) \Gamma(n_4) \Gamma(n_5) \Gamma(n_6)
    \Gamma(n_5-n_{17})}
  \nonumber \\* & \times
  \Gamma(n_{25}-n_{17}-\tfrac d2)
    \Gamma(n_{56}-n_7-\tfrac d2) \Gamma(n_{2456}-n_7-d)
  \,,
\\
\label{eq:FBE1c1cgen}
  \lefteqn{F_\BE^{\text{(1c-1c)}}(n_1,\ldots,n_7) =
    \left(\frac{M^2}{Q^2}\right)^{4-n_{13456}}
    e^{-n i\pi} \, e^{2\eps\gamma_E}}
  \nonumber \\* & \times
  \sum_{i_1,i_2\ge0}^{i_{12}\le n_7} \,
    \frac{n_7!}{i_1!\,i_2!\,(n_7-i_{12})!} \,
  \frac{\Gamma(n_1+i_2)}
    {\Gamma(n_1) \Gamma(n_3) \Gamma(n_4) \Gamma(n_5)}
  \nonumber \\ & \times
  \frac{1}{\Gamma(n_6) \Gamma(\frac d2-n_2+n_7)}
  \MBint{z_1} \MBint{z_2} \,
  \Gamma(-z_1)
  \nonumber \\ & \times
  \frac{\Gamma(n_{16}-i_1-\frac d2-z_1)
    \Gamma(n_{1356}-i_1-d-z_1)}
    {\Gamma(n_1+i_2-z_1)}
  \nonumber \\ & \times
  \Gamma(n_4+i_1+z_1)
  \Gamma(-z_2) \Gamma(\tfrac d2-n_{56}+i_1-z_2)
  \nonumber \\ & \times
  \Gamma(n_5+z_2) \,
  \frac{\Gamma(n_1-n_5+i_2-z_1-z_2)}
    {\Gamma(n_1-n_5-z_1-z_2)}
  \nonumber \\ & \times
  \Gamma(\tfrac d2-n_1+n_7-i_2+z_1+z_2)
  \nonumber \\* & \times
  \frac{\Gamma(\frac d2-n_2+n_7+z_1+z_2)}
    {\Gamma(\frac d2+n_7+z_1+z_2)}
  \,,
\\
\label{eq:FBE2c2cgen}
  \lefteqn{F_\BE^{\text{(2c-2c)}}(n_1,\ldots,n_7) =
    \left(\frac{M^2}{Q^2}\right)^{4-n_{2456}+n_7} \,
    e^{-n i\pi} \, e^{2\eps\gamma_E}}
  \nonumber \\* & \times
  \sum_{i_1,i_2,i_3\ge0}^{i_{123}\le n_7}
  \frac{n_7!}{i_1!\,i_2!\, i_3!\,(n_7-i_{123})!} \,
  \frac{\Gamma(n_1+i_1) \Gamma(n_{37}-i_1)}
    {\Gamma(n_1) \Gamma(n_2) \Gamma(n_3)}
  \nonumber \\ & \times
  \frac{\Gamma(n_2-n_{137}+i_2)}{\Gamma(n_4) \Gamma(n_5) \Gamma(n_6)} \,
  \frac{\Gamma(\tfrac d2-n_{35}+i_1) \Gamma(\frac d2-n_6+i_{23})}
    {\Gamma(\frac d2-n_{13}) \Gamma(d-n_{356}+i_{123})}
  \nonumber \\ & \times
  \Gamma(d-n_{256}+n_7+i_3) \Gamma(n_{56}-i_{23}-\tfrac d2)
  \nonumber \\* & \times
  \Gamma(n_{2456}-n_7-d)
  \,,
\\
\label{eq:FBE1c2cgen}
  \lefteqn{F_\BE^{\text{(1c-2c)}}(n_1,\ldots,n_7) =
    \left(\frac{M^2}{Q^2}\right)^{4-n_{2345}} \,
    e^{-n i\pi} \, e^{2\eps\gamma_E}}
  \nonumber \\* & \times
  \frac{\Gamma(n_2-n_{16})
    \Gamma(\frac d2-n_2) \Gamma(\frac d2-n_{56}+n_7)
    \Gamma(n_{24}-\frac d2)}
    {\Gamma(n_2) \Gamma(n_3) \Gamma(n_4)
    \Gamma(\frac d2-n_{16}) \Gamma(\frac d2-n_6+n_7)}
  \nonumber \\* & \times
  \Gamma(n_{35}-\tfrac d2)
  \,.
\end{align}
For the summation indices we use the shorthand notation
$i_{12\cdots} = i_1 + i_2 + \ldots$,
and the multiple summation is defined in the following way:
\[
  \sum_{i_1,i_2,\ldots\ge0}^{i_{12\cdots}\le n_7} =
  \sum_{i_1=0}^{n_7} \sum_{i_2=0}^{n_7-i_1} \cdots
\]

We were able to reproduce the above expressions
(\ref{eq:FBEhhgen})--(\ref{eq:FBE1c2cgen}) for the regions by writing
the full scalar integral with general $n_i$ as a triple sum over a
three-fold Mellin--Barnes integral and extracting all non-suppressed
contributions.
Our evaluation of the (h-h) region is in agreement with the known
results for the massless diagram
\cite{Kramer:1986sg,Kramer:1986sgE,Matsuura:1988sm}.

The (1c-1c) region is of order $(M^2/Q^2)^{-1}$ when
$n_1=n_3=n_4=n_5=n_6=1$. But in all these cases the inverse power of
$M^2$ is cancelled by a factor of $M^2$ in the coefficient originating
from the reduction to scalar integrals.
The purely collinear regions (1c-1c), (2c-2c) and (1c-2c) develop
poles at several points in the parameter space of the $n_i$ which need
to be regularized analytically and cancel between these three regions.

The most complicated evaluation of the contributions to the
Mercedes--Benz graph has to be performed for the (1c-1c) region with
its two-fold Mellin--Barnes integral, especially when all propagators
are present, $n_1=\cdots=n_6=1$, for $n_7=0,1,2$.
In these three cases $F_\BE$ is of order $(M^2/Q^2)^{-1}$, as
described in the previous paragraph, and only the (1c-1c) region
contributes to the leading term. In addition, the integrals are finite
with respect to both dimensional and analytic regularization, and they
result in simply a numerical constant times $(Q^2/M^2)$.
To evaluate these three complicated integrals, where none of the two
integrations can be performed explicitly due to Barnes lemmas (see
appendix~\ref{sec:MB}), we used the following strategy exemplified
here by $F_\BE(1, 1, 1, 1, 1, 1, 0)$.
After setting $d=4$ ($\eps=0$) and applying some simplifications,
equation~(\ref{eq:FBE1c1cgen}) yields
\begin{align}
  \lefteqn{F_\BE(1, 1, 1, 1, 1, 1, 0) =
    -\frac{Q^2}{M^2}
    \MBint{z_1} \MBint{z_2}}
  \nonumber \\* & \times
    \frac{\Gamma(-z_1)^2 \Gamma(z_1)
      \Gamma(-z_2)^2 \Gamma(1+z_2)
      \Gamma(1+z_1+z_2)}
      {1+z_1+z_2}
    \,,
\end{align}
where the integration contours may be chosen, e.g., as straight lines
with $\Rep z_1 = \Rep z_2 = -0.3$.
We performed the integration over $z_1$ by closing the integration
contour to the right and taking residues at the points
$z_1=0,1,2,\ldots,m,\ldots$, which are given by integrals over
$z_2$. For any given $m$, such integrals can be evaluated with
Barnes lemmas and their corollaries. We performed such calculations
up to order $m=100$. After having understood the dependence of these
integrals on~$m$, we switched to ``experimental mathematics''
(see e.g. \cite{Smirnov:2001cm} and
\cite{Fleischer:1998nb} for earlier similar examples)
and made a (successful) guess that the result of the integration over
$z_2$ can be represented in terms of nested
sums~\cite{Vermaseren:1998uu} (of the argument $m$),
in particular sign-alternating sums.
Using an ansatz as a linear combination of these nested
sums, with unknown coefficients, we solved linear systems of
equations in order to find the coefficients.
The summation of the final series, over $m$, was quite
straightforward and gave results where a value of the polylogarithm,
$\Li4(\tfrac12) \approx 0.517479$, appeared:
\begin{align}
\label{eq:FBEfullres}
  &\!\! F_\BE(1, 1, 1, 1, 1, 1, 0) =
  \frac{Q^2}{M^2} \, \biggl[
    -8\,\Li4\!\left(\frac12\right) - \frac{1}{3}\ln^4{2}
  \nonumber \\* &
    + \frac{\pi^2}{3}\ln^2{2} + \frac{19}{144}\pi^4
  \biggr] \,,
\\
  &\!\! F_\BE(1, 1, 1, 1, 1, 1, 1) =
  \frac{Q^2}{M^2} \, \biggl[
    -24\,\Li4\!\left(\frac12\right) - \ln^4{2}
  \nonumber \\* &
    + \pi^2\ln^2{2}
    + \frac{19}{48}\pi^4 - 14\zeta_3 - \pi^2 - 1
  \biggr] \,,
\\
  &\!\! F_\BE(1, 1, 1, 1, 1, 1, 2) =
  \frac{Q^2}{M^2} \, \biggl[
    -104\,\Li4\!\left(\frac12\right)
    - \frac{13}{3}\ln^4{2}
  \nonumber \\* &
    + \frac{13}{3}\pi^2\ln^2{2}
    + \frac{247}{144}\pi^4 - 63\zeta_3
    - \frac{31}{6}\pi^2 -\frac{73}{16}
  \biggr] \,.
\end{align}
We have checked these analytic constants by a direct numerical
evaluation of the Mellin--Barnes integrals.
The result~(\ref{eq:FBEfullres}) without numerator agrees
with~\cite{Aglietti:2004tq}.

The contributions from all relevant regions of all scalar integrals
sum up to the vertex correction corresponding to the Mercedes--Benz
graph in figure~\ref{fig:feynAb}c):
\begin{align}
\label{eq:BEres}
  F_{\vr,\BE} &=
    \left( C_F^2 - \frac{1}{2} C_F C_A \right)
    \left(\frac{\alpha}{4\pi}\right)^2
    \left(\frac{\mu^2}{M^2}\right)^{2\eps} S_\eps^2
    \, \Biggl\{
  \frac{1}{2\eps^2}
  \nonumber \\* & \quad
  + \frac{1}{\eps} \left[
    - \lqm^2 + 3 \lqm - \frac{2}{3}\pi^2 - \frac{13}{4} \right]
  \nonumber \\ & \quad
  + \lqm^3
  + \left(\frac{\pi^2}{3} - 7\right) \lqm^2
  + \left(8\zeta_3 - 2\pi^2 + \frac{53}{2}\right) \lqm
  \nonumber \\ & \quad
  + 128\,\Li4\!\left(\frac12\right)
  + \frac{16}{3}\ln^4{2}
  - \frac{16}{3}\pi^2\ln^2{2}
  - \frac{28}{15}\pi^4
  \nonumber \\ & \quad
  + 54\zeta_3 + \frac{115}{12}\pi^2 - \frac{263}{8}
  \Biggr\}
  + \Oc(\eps) + \Oc\!\left(\frac{M^2}{Q^2}\right)
  .
\end{align}

\subsection{Vertex correction with fermion self-energy}
\label{sec:vertexfc}

Figure~\ref{fig:feynAb}d) shows the Feynman diagram of the vertex
correction with a self-energy insertion in one of the fermion lines.
The reduction to scalar integrals as shown in
figure~\ref{fig:scalarfc} produces the following expressions:%
\begin{figure}
  \centering
  \includegraphics{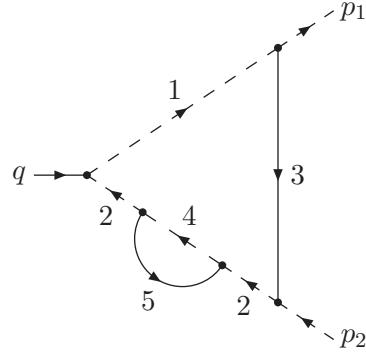}
  \caption{Scalar graph with fermion self-energy}
  \label{fig:scalarfc}
\end{figure}
\begin{align}
\label{eq:fcscalar}
  \lefteqn{F_\fc(n_1,\ldots,n_5) =
    e^{2\eps\gamma_E} \, (M^2)^{2\eps} \, (Q^2)^{n-4}}
  \nonumber \\* & \times
  \loopintf dk \loopintf d\ell \,
  \frac{1}
     {(k^2 + 2k\cdot p_1)^{n_1} \, (k^2 + 2k\cdot p_2)^{n_2}}
  \nonumber \\* & \times
  \frac{1}{(k^2 - M^2)^{n_3} \,
    ((p_2+k+\ell)^2)^{n_4} \, (\ell^2 - M^2)^{n_5}}
  \,,
\end{align}
with $k=k_3$, $\ell=k_5$ and $n = n_{12345}$.
For this graph, not every scalar product appearing in the numerator
can be expressed linearly in terms of the only five factors in the
denominator. But by applying standard tensor
reduction~\cite{Passarino:1978jh} to the subgraph of the one-loop
self-energy insertion (lines 4 and 5), the formally irreducible scalar
products may be transformed into reducible ones, so that only scalar
integrals without numerator have to be treated.

Due to the self-energy insertion the evaluation of the loop
integrations is rather easy. The complete scalar
integral~(\ref{eq:fcscalar}) with general indices~$n_i$ may be
expressed as an only two-fold Mellin--Barnes representation:
\begin{align}
  \lefteqn{F_\fc(n_1,\ldots,n_5) =
    \frac{e^{-n i\pi} \, e^{2\eps\gamma_E} \,
      \Gamma(\frac d2-n_4)}
    {\Gamma(n_1) \Gamma(n_3) \Gamma(n_4) \Gamma(n_5)}}
  \nonumber \\* & \times
  \MBint{z_1} \MBint{z_2} \,
  \left(\frac{M^2}{Q^2}\right)^{2\eps+z_1} \,
  \frac{\Gamma(d-n_{2345}-z_1)}{\Gamma(\frac32d-n_{12345}-z_1)}
  \nonumber \\ & \times
  \Gamma(n_{12345}-d+z_1) \,
  \Gamma(-z_2) \Gamma(n_3+z_2) \Gamma(\tfrac d2-n_1+z_2)
  \nonumber \\ & \times
  \frac{\Gamma(-n_3-z_1-z_2) \Gamma(\frac d2-n_{35}-z_1-z_2)}
    {\Gamma(d-n_{345}-z_1-z_2)}
  \nonumber \\* & \times
  \frac{\Gamma(n_{345}-\frac d2+z_1+z_2)}
    {\Gamma(n_{2345}-\frac d2+z_1+z_2)}
  \,.
\end{align}
{}From this expression, the residues producing non-sup\-pressed
contributions are extracted. They correspond exactly to the
contributions from the following five regions:
\begin{alignat*}{6}
  \text{(h-h):} & \quad && k \sim Q, \quad && \ell \sim Q \\
  \text{(1c-h):} &&& k \;\|\; p_1, && \ell \sim Q \\
  \text{(2c-2c):} &&& k \;\|\; p_2, && \ell \;\|\; p_2 \\
  \text{(h-s):} &&& k \sim Q, && \ell \sim M \\
  \text{(1c-s):} &&& k \;\|\; p_1, && \ell \sim M
\end{alignat*}
These contributions are evaluated as
\begin{align}
\label{eq:Ffchhgen}
  \lefteqn{F_\fc^{\text{(h-h)}}(n_1,\ldots,n_5) =
    \left(\frac{M^2}{Q^2}\right)^{2\eps} \,
    e^{-n i\pi} \, e^{2\eps\gamma_E}}
  \nonumber \\* & \times
  \frac{\Gamma(\frac d2-n_{13}) \Gamma(\frac d2-n_4) \Gamma(\frac d2-n_5)
    \Gamma(d-n_{2345})}
    {\Gamma(n_1) \Gamma(n_4) \Gamma(n_5)
    \Gamma(d-n_{45}) \Gamma(\frac32d-n_{12345})}
  \nonumber \\* & \times
  \frac{\Gamma(n_{45}-\frac d2) \Gamma(n_{12345}-d)}
    {\Gamma(n_{245}-\frac d2)}
  \,,
\\
\label{eq:Ffc1chgen}
  \lefteqn{F_\fc^{\text{(1c-h)}}(n_1,\ldots,n_5) =
  \left(\frac{M^2}{Q^2}\right)^{2-n_{13}+\eps} \,
  e^{-n i\pi} \, e^{2\eps\gamma_E}}
  \nonumber \\* & \times
  \frac{\Gamma(\frac d2-n_1) \Gamma(\frac d2-n_4) \Gamma(\frac d2-n_5)
    \Gamma(\frac d2+n_1-n_{245})}
    {\Gamma(n_1) \Gamma(n_3) \Gamma(n_4) \Gamma(n_5)
    \Gamma(d-n_{45}) \Gamma(d-n_{245})}
  \nonumber \\* & \times
  \Gamma(n_{13}-\tfrac d2) \Gamma(n_{45}-\tfrac d2)
  \,,
\\
\label{eq:Ffc2c2cgen}
  \lefteqn{F_\fc^{\text{(2c-2c)}}(n_1,\ldots,n_5) =
    \left(\frac{M^2}{Q^2}\right)^{4-n_{2345}} \,
    e^{-n i\pi} \, e^{2\eps\gamma_E}}
  \nonumber \\* & \times
  \frac{\Gamma(\frac d2-n_4)}
    {\Gamma(n_3) \Gamma(n_4) \Gamma(n_5) \Gamma(\frac d2-n_1)}
  \MBint z \,
  \frac{\Gamma(-z)}{\Gamma(n_2-z)}
  \nonumber \\ & \times
  \Gamma(n_{24}-\tfrac d2-z) \Gamma(n_{245}-d-z) \,
  \Gamma(n_3+z)
  \nonumber \\* & \times
  \frac{\Gamma(\frac d2-n_1+z) \Gamma(\frac d2-n_2+z)}
    {\Gamma(\frac d2+z)}
  \,,
\\
\label{eq:Ffchsgen}
  \lefteqn{F_\fc^{\text{(h-s)}}(n_1,\ldots,n_5) =
    \left(\frac{M^2}{Q^2}\right)^{2-n_5+\eps} \,
    e^{-n i\pi} \, e^{2\eps\gamma_E}}
  \nonumber \\* & \times
  \frac{\Gamma(\frac d2-n_{13}) \Gamma(\frac d2-n_{234})
    \Gamma(n_{1234}-\frac d2) \Gamma(n_5-\frac d2)}
    {\Gamma(n_1) \Gamma(n_5) \Gamma(n_{24})
    \Gamma(d-n_{1234})}
  \,,
\\
\label{eq:Ffc1csgen}
  \lefteqn{F_\fc^{\text{(1c-s)}}(n_1,\ldots,n_5) =
    \left(\frac{M^2}{Q^2}\right)^{4-n_{135}} \,
    e^{-n i\pi} \, e^{2\eps\gamma_E}}
  \nonumber \\* & \times
  \frac{\Gamma(n_1-n_{24}) \Gamma(\frac d2-n_1)
    \Gamma(n_{13}-\frac d2) \Gamma(n_5-\frac d2)}
    {\Gamma(n_1) \Gamma(n_3) \Gamma(n_5)
    \Gamma(\frac d2-n_{24})}
  \,.
\end{align}
The contribution of the (h-h) region is known from the massless
diagram~\cite{Kramer:1986sg,Kramer:1986sgE,Matsuura:1988sm,Gonsalves:1983nq}.
In the reduction to scalar integrals only parameters~$n_i$ with
$n_2\le2$ and $n_i\le1$, $i=1,3,4,5$, are involved. Therefore the
contributions of the (h-s) and (1c-s) regions are always
suppressed by at least one factor $M^2/Q^2$.
On the other hand, the (2c-2c) region is of order $(M^2/Q^2)^{-1}$ if
$n_2=2$, $n_3=n_4=n_5=1$, but this inverse power of $M^2$ is cancelled
by a factor of $M^2$ from the reduction to scalar integrals.

For the leading order in $M^2/Q^2$, no analytic regularization is
necessary.
The contributions of the (h-h), (1c-h) and (2c-2c) regions sum up to
the results for the scalar integrals, e.g.
\begin{align}
\label{eq:Ffcfullres}
  F_\fc(1, 1, 1, 1, 1) &=
  \frac{1}{\eps} \left(-\frac{1}{2}\lqm^2 - \frac{\pi^2}{3}\right)
  + \frac{1}{2}\lqm^3 - \lqm^2
  \nonumber \\* & \qquad
  + 4\zeta_3 - \frac{\pi^2}{3} \,,
\\
  F_\fc(1, 2, 1, 1, 1) &=
  \frac{Q^2}{M^2} \, \biggl[
  -\frac{1}{\eps^2} - \frac{1}{\eps} - \frac{\pi^2}{3} - \frac{3}{2}
  \biggr] \,.
\end{align}
The whole vertex correction originating from the Feynman diagram in
figure~\ref{fig:feynAb}d) evaluates to
\begin{align}
\label{eq:fcres}
  F_{\vr,\fc} &=
    C_F^2
    \left(\frac{\alpha}{4\pi}\right)^2
    \left(\frac{\mu^2}{M^2}\right)^{2\eps} S_\eps^2
    \, \Biggl\{
  -\frac{1}{2\eps^2}
  \nonumber \\* & \qquad
  + \frac{1}{\eps} \left[
    \lqm^2 - 3 \lqm + \frac{2}{3}\pi^2 + \frac{13}{4} \right]
  - \lqm^3
  + 5 \lqm^2
  \nonumber \\ & \qquad
  - \frac{33}{2} \lqm
  - 8\zeta_3 - \frac{\pi^2}{4} + \frac{171}{8}
  \Biggr\}
  + \Oc(\eps) + \Oc\!\left(\frac{M^2}{Q^2}\right)
  .
\end{align}

\subsection{Vertex correction with non-Abelian Mercedes--Benz graph}
\label{sec:vertexBECA}

\begin{figure}
  \centering
  \includegraphics{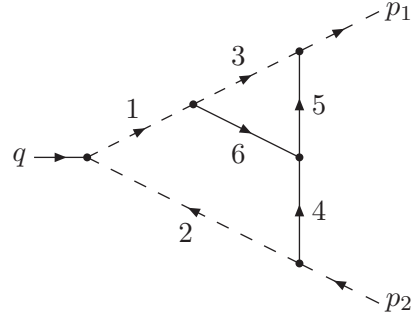}
  \caption{Scalar non-Abelian Mercedes--Benz graph}
  \label{fig:scalarBECA}
\end{figure}

The Mercedes--Benz graph in figure~\ref{fig:feynNA}a) is of pure
non-Abelian nature due to its three-gauge-boson vertex.
The corresponding scalar integrals are illustrated in
figure~\ref{fig:scalarBECA}%
and defined as follows:
\begin{align}
\label{eq:BECAscalar}
  \lefteqn{F_\BECA(n_1,\ldots,n_7) =
    e^{2\eps\gamma_E} \, (M^2)^{2\eps} \, (Q^2)^{n-n_7-4}}
  \nonumber \\* & \times
  \loopintf dk \loopintf d\ell \,
  \frac{(2p_2 \cdot k)^{n_7}}
    {(\ell^2-2p_1\cdot\ell)^{n_1} \, (\ell^2-2p_2\cdot\ell)^{n_2}}
  \nonumber \\ & \times
  \frac{1}{(k^2 - 2p_1\cdot k)^{n_3} \,
    (\ell^2 - M^2)^{n_4} \, (k^2-M^2)^{n_5}}
  \nonumber \\* & \times
  \frac{1}{((k-\ell)^2-M^2)^{n_6}} \,,
\end{align}
with $k=k_5$, $\ell=k_4$ and $n=n_{123456}$.
This definition is the same as for the Abelian Mercedes--Benz graph in
figure~\ref{fig:scalarBE} and equation~(\ref{eq:BEscalar}),
except for the distribution of the masses in the propagators.
Also the list of relevant regions is similar, but now there is the
(s'-h) region instead of the (us-2c) region:
\begin{alignat*}{6}
  \text{(h-h):} & \quad && k \sim Q, \quad && \ell \sim Q \\
  \text{(1c-h):} &&& k \;\|\; p_1, && \ell \sim Q \\
  \text{(h-2c):} &&& k \sim Q, && \ell \;\|\; p_2 \\
  \text{(s'-h):} &&& k_6 \sim M, \quad && \ell \sim Q \\
  \text{(1c-1c):} &&& k \;\|\; p_1, && \ell \;\|\; p_1 \\
  \text{(2c-2c):} &&& k \;\|\; p_2, && \ell \;\|\; p_2 \\
  \text{(1c-2c):} &&& k \;\|\; p_1, && \ell \;\|\; p_2
\end{alignat*}
The (s'-h) region is of order $(M^2/Q^2)^{2-n_6+\eps}$ and therefore
suppressed with respect to the (h-h) region, as $n_i\le1$
($i=1,\ldots,6$).
The (2c-2c) region is of order $(M^2/Q^2)^{4-n_{2456}+n_7}$ and
suppressed for $n_7>0$; it is only evaluated for $n_7=0$.

The leading contributions of regions with $k_3, k_5, k_6 \sim Q$ are
identical to the corresponding contributions of the Abelian
Mercedes--Benz graph:
\begin{align}
  F_\BECA^{\text{(h-h)}}(n_1,\ldots,n_7) &=
    F_\BE^{\text{(h-h)}}(n_1,\ldots,n_7) \,,
\\
  F_\BECA^{\text{(h-2c)}}(n_1,\ldots,n_7) &=
    F_\BE^{\text{(h-2c)}}(n_1,\ldots,n_7) \,.
\end{align}
The contributions of the other regions are given by
\begin{align}
\label{eq:FBECA1chgen}
  \lefteqn{F_\BECA^{\text{(1c-h)}}(n_1,\ldots,n_7) =
    \left(\frac{M^2}{Q^2}\right)^{2-n_{35}+\eps} \,
    e^{-n i\pi} \, e^{2\eps\gamma_E}}
  \nonumber \\* & \times
  \frac{\Gamma(\frac d2-n_3) \Gamma(\frac d2-n_{146}) \Gamma(n_{35}-\frac d2)}
    {\Gamma(n_1) \Gamma(n_2) \Gamma(n_3) \Gamma(n_5) \Gamma(n_6)
    \Gamma(d-n_{1246})}
  \nonumber \\ & \times
  \MBint z \,
  \frac{\Gamma(-z) \Gamma(\frac d2-n_{246}-z)
    \Gamma(n_6+z) \Gamma(n_{37}+z)}
    {\Gamma(\frac d2+n_7+z)}
  \nonumber \\* & \times
  \Gamma(n_{1246}-\tfrac d2+z)
  \,,
\\
\label{eq:FBECA1c1cgen}
  \lefteqn{F_\BECA^{\text{(1c-1c)}}(n_1,\ldots,n_7) =
    \left(\frac{M^2}{Q^2}\right)^{4-n_{13456}} \,
    e^{-n i\pi} \, e^{2\eps\gamma_E}}
  \nonumber \\* & \times
  \sum_{i_1,i_2,i_3\ge0}^{i_{123}\le n_7}
  \frac{n_7!}{i_1!\,i_2!\,i_3!\,(n_7-i_{123})!} \,
  \frac{\Gamma(n_{17}-i_{12})}
    {\Gamma(n_1) \Gamma(n_3) \Gamma(n_4) \Gamma(n_5)}
  \nonumber \\ & \times
  \frac{1}{\Gamma(n_6) \Gamma(\frac d2-n_2+n_7)}
  \MBint{z_1} \MBint{z_2} \,
  \Gamma(-z_1)
  \Gamma(-z_2)
  \nonumber \\ & \times
  \Gamma(n_5+z_1) \Gamma(\tfrac d2-n_3+n_7-i_{123}+z_1)
  \Gamma(n_4+i_1+z_2)
  \nonumber \\ & \times
  \frac{\Gamma(n_{1467}-i_{123}-\tfrac d2+z_2)
    \Gamma(n_{13456}-d+z_2)}
    {\Gamma(n_{147}-i_2+z_2)}
  \nonumber \\ & \times
  \frac{\Gamma(n_{134}+i_{13}-\tfrac d2-z_1+z_2)}
    {\Gamma(n_{14567}-i_{123}-\frac d2+z_1+z_2)}
  \nonumber \\ & \times
  \frac{\Gamma(\frac d2-n_{14}+i_2+z_1-z_2)}
    {\Gamma(\frac d2-n_4+n_7-i_1+z_1-z_2)}
  \nonumber \\* & \times
  \Gamma(\tfrac d2-n_{24}+n_7-i_1+z_1-z_2)
  \,,
\\
\label{eq:FBECA2c2cgen}
  \lefteqn{F_\BECA^{\text{(2c-2c)}}(n_1,\ldots,n_6,n_7=0) =
    \left(\frac{M^2}{Q^2}\right)^{4-n_{2456}} \,
    e^{-n i\pi} \, e^{2\eps\gamma_E}}
  \nonumber \\* & \times
  \frac{\Gamma(n_2-n_{13})}
    {\Gamma(n_2) \Gamma(n_4) \Gamma(n_5) \Gamma(n_6) \Gamma(\frac d2-n_{13})}
  \MBint z \,
  \Gamma(-z)
  \nonumber \\ & \times
  \frac{\Gamma(n_{24}-\frac d2-z)
    \Gamma(n_5+z) \Gamma(n_6-n_3+z)}
    {\Gamma(n_{56}-n_3+2z)}
  \nonumber \\* & \times
  \Gamma(\tfrac d2-n_2+z) \Gamma(n_{56}-\tfrac d2+z)
  \,,
\\
\label{eq:FBECA1c2cgen}
  \lefteqn{F_\BECA^{\text{(1c-2c)}}(n_1,\ldots,n_7) =
    \left(\frac{M^2}{Q^2}\right)^{4-n_{2345}} \,
    e^{-n i\pi} \, e^{2\eps\gamma_E}}
  \nonumber \\* & \times
  \frac{\Gamma(n_2-n_{16}) \Gamma(n_{37}-n_6)
    \Gamma(\frac d2-n_2) \Gamma(\frac d2-n_3)}
    {\Gamma(n_2) \Gamma(n_3) \Gamma(n_4) \Gamma(n_5)
    \Gamma(\frac d2-n_{16}) \Gamma(\frac d2-n_6+n_7)}
  \nonumber \\* & \times
  \Gamma(n_{24}-\tfrac d2) \Gamma(n_{35}-\tfrac d2)
  \,.
\end{align}
The evaluation of this non-Abelian vertex graph is more complicated
than in the Abelian case, mainly due to the appearance of three
massive propagators. The complete summation of the infinite number of
residues in the Mellin--Barnes integrals (\ref{eq:FBECA1c1cgen}) and
(\ref{eq:FBECA2c2cgen}) is quite intricate.
As the calculation of the four-fermion amplitude demands the result of
the form factor only to N$^3$LL accuracy, we have refrained from
calculating the non-logarithmic constant in the non-Abelian
corrections (cf. the beginning of section~\ref{sec:vertex}).
Therefore we have only extracted all logarithms $\ln(Q^2/M^2)$ from
the integrals.

The (1c-h) region has been evaluated in the usual way as described in
the previous sections.
{}From the (c-c) regions, i.e. (1c-1c), (2c-2c) and (1c-2c), the logarithmic
contributions have been isolated. The expressions for the regions
depend on $M^2/Q^2$ only through a prefactor of the form
$(M^2/Q^2)^{m+x}$, where $m$ is an integer and $x$ is made up of
regularization parameters (like~$\eps$) tending to zero.
Logarithms $\ln(Q^2/M^2)$ arise only when poles in the regularization
parameters appear, e.g.
\[
  \left(\frac{M^2}{Q^2}\right)^x \, \frac{1}{x} =
  \frac{1}{x} - \ln\!\left(\frac{Q^2}{M^2}\right) + \Oc(x) \,.
\]
As the exponents of these prefactors in the contributions
(\ref{eq:FBECA1c1cgen})--(\ref{eq:FBECA1c2cgen}) of the (c-c) regions
involve only the parameters~$n_i$ and not $\eps$, only poles
originating from the analytic regularization may give rise to logarithms.
A thorough analysis of the Mellin--Barnes integrals shows that such
poles only appear in the following seven integrals:
\begin{align*}
  & F_\BECA^{\text(c-c)}(-1,1,1,1,1,1,0) \,, \\
  & F_\BECA^{\text(c-c)}(0,1,1,1,1,1,n_7) \,\quad \text{with $n_7=0,1,2$}, \\
  & F_\BECA^{\text(c-c)}(1,1,0,1,1,1,0) \,\quad \text{and} \\
  & F_\BECA^{\text(c-c)}(1,1,1,1,1,0,n_7) \,\quad \text{with $n_7=0,1$}.
\end{align*}
When closing the integration contours in the Mellin--Barnes integrals,
it is sufficient to take those residues which are responsable for the
poles. In most cases these are only a finite number of residues.
Only the integrals $F_\BECA^{\text(c-c)}(0,1,1,1,1,1,0)$ and
$F_\BECA^{\text(c-c)}(1,1,0,1,1,1,0)$ need the summation of an
infinite number of residues.
Such summations can be transformed to infinite series like
\begin{multline}
\label{eq:suminvbinomS1}
  \sum_{m=0}^\infty \frac{1}{\binom{2m}{m}}
    \left(\frac{1}{3+2m} - \frac{1}{1+2m}\right)
  \\* \times
    \left(\frac{1}{3+2m} + \frac{1}{1+2m} + S_1(2m) - S_1(m)\right) ,
\end{multline}
where $S_1(m) = \sum_{i=1}^m \frac{1}{i}$ is a harmonic sum.
Recently there has been a lot of progess in solving summations
like~(\ref{eq:suminvbinomS1}) analytically (see
e.g. \cite{Moch:2001zr,Weinzierl:2004bn}). But still not all possible
cases are covered, and the transformation of a given expression into a
series where the solution is known can be quite cumbersome.

On the other hand, the series in (\ref{eq:suminvbinomS1}) is
converging very fast. The numerical summation of the first 300 terms
approximates the series with an accuracy of more than 100 decimal
digits.
This enables us to use the following method (see
e.g. \cite{Kalmykov:2000qe}).
An ansatz is chosen as a linear combination of analytical constants
like $\pi^2$, $\zeta_3$, $\ln^4 2$ etc. with unknown rational
coefficients. The determination of the coefficients starting from the
numerical result is performed by the PSLQ
algorithm~\cite{Ferguson:1991,Bailey:1993,Ferguson:1996}.
We have used an implementation~\cite{Veretin:PSLQ} of PSLQ in Fortran
with multiprecision arithmetic~\cite{Bailey:1990,Bailey:1991}.
The series above in~(\ref{eq:suminvbinomS1}) has hereby been
identified with the analytical expression
$4\sqrt{3}\,\Cl2(\tfrac{\pi}{3}) - 8$,
where $\Cl2(\frac{\pi}{3}) \approx 1.014942$ is a value of the Clausen
function.

In addition to the logarithms, we have calculated in a purely
analytical way the complete set of poles in~$\eps$ in order to control
the cancellation of ultraviolet and infrared singularities.

The contributions from all regions sum up to the results of the scalar
integrals, from which we quote the two most complicated ones:
\begin{align}
  F_\BECA(0, 1, 1, 1, 1, 1, 0) &=
  -\frac{1}{12}\lqm^4 - \frac{\pi^2}{6}\lqm^2 + \frac{2}{3}\zeta_3\lqm \,,
\\
  F_\BECA(1, 1, 0, 1, 1, 1, 0) &=
  \frac{1}{\eps} \left(-\frac{1}{2}\lqm^2 - \frac{\pi^2}{3}\right)
  + \frac{1}{3}\lqm^3 - \lqm^2
  \nonumber \\* & \qquad
  + 2\sqrt{3}\Cl2\!\left(\frac{\pi}{3}\right)\lqm \,,
\end{align}
where non-logarithmic terms of order $\eps^0$ have been omitted.
The integrals $F_\BECA(1,n_2,1,1,1,1,n_7)$ are of order
$(M^2/Q^2)^{-1}$, with the only contribution coming from the (1c-1c)
region, and the inverse power of $M^2$ is again cancelled by a factor
of $M^2$ from the reduction to scalar integrals. These integrals,
however, produce neither logarithms $\lqm=\ln(Q^2/M^2)$ nor poles in
$\eps$ and do therefore not contribute to the result in N$^3$LL accuracy.
The result for the vertex correction corresponding to the non-Abelian
Mercedes--Benz graph in figure~\ref{fig:feynNA}a) is as follows:
\begin{align}
\label{eq:BECAres}
  F_{\vr,\BECA} &=
    C_F C_A
    \left(\frac{\alpha}{4\pi}\right)^2
    \left(\frac{\mu^2}{M^2}\right)^{2\eps} S_\eps^2
    \, \Biggl\{
  \frac{3}{4\eps^2}
  \nonumber \\* & \qquad
  + \frac{1}{\eps} \left[
    - \frac{3}{2} \lqm^2 + \frac{9}{2} \lqm - \pi^2 - \frac{37}{8} \right]
  + \frac{1}{12} \lqm^4
  \nonumber \\ & \qquad
  + \frac{1}{2} \lqm^3
  + \left(\frac{\pi^2}{6} - \frac{11}{2}\right) \lqm^2
  \nonumber \\ & \qquad
  + \left(4\sqrt{3}\,\Cl2\!\left(\frac{\pi}{3}\right)
    - \frac{2}{3}\zeta_3 - \frac{5}{6}\pi^2 + \frac{89}{4}\right) \lqm
  \Biggr\}
  \nonumber \\* & \quad
  + \Oc(\eps^0 \lqm^0) + \Oc(\eps) + \Oc\!\left(\frac{M^2}{Q^2}\right)
  .
\end{align}

\subsection{Non-Abelian vertex corrections with loop insertions}
\label{sec:vertexWc}

This section treats all vertex diagrams from figure~\ref{fig:feynNA}
where a self-energy loop has been inserted in the gauge boson
propagator: the gauge boson loop in figure~\ref{fig:feynNA}b),
the ghost field loop in figure~\ref{fig:feynNA}c)
and loops involving the Higgs and Goldstone bosons in figures
\ref{fig:feynNA}d), \ref{fig:feynNA}e) and \ref{fig:feynNA}f).
Care must be taken in the interpretation of the Feynman rules
(appendix~\ref{sec:feynman}) not to forget the factor $(-1)$ for the
loop of the anticommuting ghost fields and the symmetry factor $1/2$
for the loops with two gauge bosons or two Goldstone bosons.

Additional contributions from ``tadpoles'', where a loop of only one
gauge, Higgs or Goldstone boson is attached to the gauge boson
propagator via a vertex with four fields, are omitted here because
they are cancelled exactly by the corresponding contributions from the
renormalization of the gauge boson mass (see section~\ref{sec:ren}).

For the Higgs boson mass~$M_H$ we use the approximation $M_H = M_W$,
which facilitates the loop calculations. The form factor depends on
the Higgs mass only in N$^3$LL accuracy, i.e. via the coefficient of
the linear logarithm, and the higher powers of the electroweak
logarithm are not affected by changes in the Higgs mass.
We have checked explicitly by evaluating the Higgs contributions for
the hypothetical case $M_H = 0$ (see the discussion of the results in
section~\ref{sec:result}) that effects due to a wrong value of the
Higgs mass are indeed negligible.

\begin{figure}
  \centering
  \includegraphics{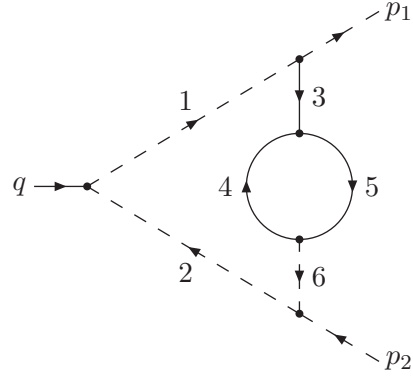}
  \caption{Scalar graph for non-Abelian vertex corrections with loop
    insertions}
  \label{fig:scalarWc}
\end{figure}

As in our approximation (and using the Feynman--'t~Hooft gauge) all
particles running in the self-energy loop have the same mass $M=M_W$,
the vertex corrections of this section share the same set of scalar
integrals, which are illustrated in figure~\ref{fig:scalarWc}%
and defined in the following equation:
\begin{align}
\label{eq:Wcorrscalar}
  \lefteqn{F_\Wc(n_1,\ldots,n_6) =
    e^{2\eps\gamma_E} \, (M^2)^{2\eps} \, (Q^2)^{n-4}}
  \nonumber \\* & \times
  \loopintf dk \loopintf d\ell \,
  \frac{1}
    {(k^2+2p_1\cdot k)^{n_1} \, (k^2+2p_2\cdot k)^{n_2}}
  \nonumber \\ & \times
  \frac{1}{(k^2 - M^2)^{n_3} \, (\ell^2 - M^2)^{n_4} \,
    ((k+\ell)^2 - M^2)^{n_5} \, (k^2)^{n_6}}
  \,,
\end{align}
with $k=k_3$ and $\ell=k_4$.
The additional massless propagator corresponding to the
parameter~$n_6$ is introduced when performing a tensor reduction on
the self-energy loop (lines 4 and 5) in order to eliminate the scalar
products in the numerator of the integral.
The presence of both a massive and a massless propagator with the same
momentum $k_3 = k_6 = k$ could, of course, be avoided by partial
fractioning. But this would produce factors of $1/M^2$, complicating
the expansion in $M^2/Q^2$. So we remained with both propagators in
the scalar integrals~(\ref{eq:Wcorrscalar}). In order to avoid
ambiguities in the reduction to scalar integrals, we fixed $n_3=2$ and
cancelled factors of~$k^2$ in the numerator exclusively with the sixth
propagator.

In general the following regions are relevant:
\begin{alignat*}{6}
  \text{(h-h):} & \quad && k \sim Q, \quad && \ell \sim Q \\
  \text{(h-s):} &&& k \sim Q, && \ell \sim M \\
  \text{(h-s'):} &&& k \sim Q, && k_5 \sim M \\
  \text{(1c-1c):} &&& k \;\|\; p_1, && \ell \;\|\; p_1 \\
  \text{(2c-2c):} &&& k \;\|\; p_2, && \ell \;\|\; p_2
\end{alignat*}
But since the (h-s) regions is of order $(M^2/Q^2)^{2-n_4+\eps}$ and
the (h-s') regions is of order $(M^2/Q^2)^{2-n_5+\eps}$, they are both
suppressed with respect to the (h-h) region for all relevant cases.
The contributions from the other regions can be expressed as follows:
\begin{align}
\label{eq:FWchhgen}
  \lefteqn{F_\Wc^{\text{(h-h)}}(n_1,\ldots,n_6) =
    \left(\frac{M^2}{Q^2}\right)^{2\eps} \,
    e^{-n i\pi} \, e^{2\eps\gamma_E}}
  \nonumber \\* & \times
  \frac{\Gamma(\frac d2-n_4) \Gamma(\frac d2-n_5)
    \Gamma(d-n_{13456}) \Gamma(d-n_{23456})}
    {\Gamma(n_1) \Gamma(n_2) \Gamma(n_4) \Gamma(n_5)
    \Gamma(d-n_{45})
    \Gamma(\frac32d-n_{123456})}
  \nonumber \\* & \times
  \Gamma(n_{45}-\tfrac d2) \Gamma(n_{123456}-d)
  \,,
\\
\label{eq:FWc1c1cgen}
  \lefteqn{F_\Wc^{\text{(1c-1c)}}(n_1,\ldots,n_6) =
    \left(\frac{M^2}{Q^2}\right)^{4-n_{13456}} \,
    e^{-n i\pi} \, e^{2\eps\gamma_E}}
  \nonumber \\* & \times
  \frac{\Gamma(n_1-n_2)}
    {\Gamma(n_1) \Gamma(n_3) \Gamma(n_4) \Gamma(n_5)
    \Gamma(\frac d2-n_2)} \,
  \MBint z \,
  \Gamma(-z)
  \nonumber \\* & \times
  \frac{\Gamma(n_{136}-\frac d2-z)
    \Gamma(n_4+z) \Gamma(n_5+z)
    \Gamma(\frac d2-n_{16}+z)}
    {\Gamma(n_{45}+2z)}
  \nonumber \\* & \times
  \Gamma(n_{45}-\tfrac d2+z)
  \,.
\end{align}
For symmetry reasons, $F_\Wc^{\text{(2c-2c)}}$ can be obtained from
$F_\Wc^{\text{(1c-1c)}}$ by exchanging $n_1 \leftrightarrow n_2$.
Some of the (1c-1c) and (2c-2c) contributions are of order
$(M^2/Q^2)^{-1}$, but this inverse power of~$M^2$ is always cancelled
by a factor of~$M^2$ originating either from the reduction to scalar
integrals or from the Feynman rules, when two $WWH$-vertices are
present.

As for the non-Abelian Mercedes--Benz graph in the previous section, we
have only extracted the logarithms and the poles in~$\eps$.
The (c-c) regions (1c-1c) and (2c-2c) produce logarithmic terms for
$n_1 = n_2 = 1$. For most of the (c-c) contributions, the evaluation
of a finite number of residues in the complex $z$-plane is
sufficient. Only the two cases $F_\Wc^{\text{(c-c)}}(1,1,2,1,1,0)$ and
$F_\Wc^{\text{(c-c)}}(1,1,2,1,1,-1)$ demand the summation of an
infinite number of residues. We have solved these two summations
numerically and found the corresponding analytic expressions with the
help of the PSLQ algorithm.
In all cases the extraction of the poles in~$\eps$ required only a
finite number of residues.

We quote the results for the two most complicated scalar integrals:
\begin{align}
  &\!\! F_\Wc(1, 1, 2, 1, 1, 0) = \frac{Q^2}{M^2} \, \biggl[
    \frac{1}{\eps}
    - \frac{4}{3}\sqrt{3}\Cl2\!\left(\frac{\pi}{3}\right)
    + 2
  \biggr] \, \lqm \,,
\\
  &\!\! F_\Wc(1, 1, 2, 1, 1, -1) =
    \frac{1}{\eps} \left(-\frac{1}{2}\lqm^2 + \lqm - \frac{\pi^2}{3}\right)
    + \frac{1}{3}\lqm^3
  \nonumber \\* &
    - \lqm^2
    + \left(\frac{2}{3}\sqrt{3}\Cl2\!\left(\frac{\pi}{3}\right) + 2\right)
      \lqm
  \,.
\end{align}
The vertex corrections corresponding to the Feynman diagrams in figures
\ref{fig:feynNA}b)--\ref{fig:feynNA}f) are obtained by inserting the
results for the scalar integrals into the expressions returned from
the reduction of each diagram.

The vertex corrections with the non-Abelian gauge boson and ghost
field loops, figures \ref{fig:feynNA}b) and \ref{fig:feynNA}c), have
been evaluated together. Their sum is
\begin{align}
\label{eq:WWccres}
  F_{\vr,\WWcc} &=
    C_F C_A
    \left(\frac{\alpha}{4\pi}\right)^2
    \left(\frac{\mu^2}{M^2}\right)^{2\eps} S_\eps^2
    \, \Biggl\{
  \nonumber \\* & \qquad
  \frac{1}{\eps} \left[
    - \frac{5}{3} \lqm^2 + \frac{49}{3} \lqm - \frac{10}{9}\pi^2
    - \frac{337}{12} \right]
  + \frac{10}{9} \lqm^3
  \nonumber \\ & \qquad
  - \frac{76}{9} \lqm^2
  + \left(-4\sqrt{3}\,\Cl2\!\left(\frac{\pi}{3}\right)
    + \frac{859}{18}\right) \lqm
  \Biggr\}
  \nonumber \\* & \quad
  + \Oc(\eps^0 \lqm^0) + \Oc(\eps) + \Oc\!\left(\frac{M^2}{Q^2}\right)
  .
\end{align}
The vertex correction in figure~\ref{fig:feynNA}d) with gauge and
Higgs boson in the loop insertion contains two factors of~$M$ from the
two $WWH$-vertices, but these are cancelled by factors $1/M^2$ in the
results of some of the scalar integrals. So this vertex correction is
not suppressed:
\begin{align}
\label{eq:WHvertexres}
  F_{\vr,\WH} &=
    \left(\frac{\alpha}{4\pi}\right)^2
    \left(\frac{\mu^2}{M^2}\right)^{2\eps} S_\eps^2
    \, \Biggl\{
  \frac{1}{\eps} \left[ -\frac{3}{2} \lqm + 3 \right]
  \nonumber \\* & \qquad
  + \biggl( 2\sqrt{3}\,\Cl2\!\left(\frac{\pi}{3}\right)
    - 3 \biggr) \, \lqm
  \Biggr\}
  \nonumber \\* & \quad
  + \Oc(\eps^0 \lqm^0) + \Oc(\eps) + \Oc\!\left(\frac{M^2}{Q^2}\right)
  .
\end{align}
The two vertex corrections in figures \ref{fig:feynNA}e) and
\ref{fig:feynNA}f) with Higgs and Goldstone bosons in the loop
insertion yield the same result (for $M_H=M$):
\begin{align}
\label{eq:Hphiphivertexres}
  F_{\vr,\Hphi} &= F_{\vr,\phiphi} =
    \left(\frac{\alpha}{4\pi}\right)^2
    \left(\frac{\mu^2}{M^2}\right)^{2\eps} S_\eps^2
    \, \Biggl\{
  \nonumber \\* & \qquad
  \frac{1}{\eps} \left[
    \frac{1}{16} \lqm^2 + \frac{7}{16} \lqm
     + \frac{\pi^2}{24} - \frac{67}{64} \right]
  - \frac{1}{24} \lqm^3
  \nonumber \\ & \qquad
  + \frac{17}{48} \lqm^2
  + \biggl( -\frac{3}{4}\sqrt{3}\,\Cl2\!\left(\frac{\pi}{3}\right)
    + \frac{19}{96} \biggr) \, \lqm
  \Biggr\}
  \nonumber \\* & \quad
  + \Oc(\eps^0 \lqm^0) + \Oc(\eps) + \Oc\!\left(\frac{M^2}{Q^2}\right)
  .
\end{align}
The contributions involving Higgs and Goldstone bosons are only valid
for the spontaneously broken $SU(2)$ model, they cannot be transformed
e.g. to a $U(1)$ model simply by setting other values for $C_F$, $C_A$
and $T_F$. We have therefore written these contributions with the
Casimir operators already replaced by their $SU(2)$ values.

\section{Renormalization contributions}
\label{sec:ren}

Section~\ref{sec:vertex} has treated the evaluation of the vertex
corrections which contribute to the Abelian vector form factor.
These have been performed with Feynman rules originating from the
unrenormalized Lagrangian. Therefore the contributions due to the
renormalization of the fields (section~\ref{sec:renfield}), the
coupling constant (section~\ref{sec:renalpha}) and the gauge boson
mass (section~\ref{sec:renmass}) have to be added.

\subsection{Field renormalization}
\label{sec:renfield}

The renormalization of the two fermion fields in the Abelian vector
current requires the multiplication of the vertex corrections by a
factor of $Z_f$, where $(Z_f)^{1/2}$ is the fermion field
renormalization constant.
On the other hand, $Z_f$ is determined by the fermion self-energy
corrections~$\Sigma$ at on-shell momentum $p^2=0$ (for massless
fermions).
In a perturbative expansion, the field renormalization constant is
$Z_f = 1 + \Sigma_1 + \Sigma_2 + \Oc(\alpha^3)$
and the vertex corrections are
$F_\vr = 1 + F_{\vr,1} + F_{\vr,2} + \Oc(\alpha^3)$,
where the indices 1 and 2 indicate the one- and two-loop
contributions, respectively.

The total Abelian vector form factor up to order~$\alpha^2$ can be
written as
\begin{align*}
  F &= F_\vr \cdot Z_f
  \nonumber \\* &=
  1 +{} \underbrace{F_{\vr,1} + \Sigma_1}_{\Oc(\alpha)}
  {}+{} \underbrace{F_{\vr,2} + \Sigma_2 + F_{\vr,1} \Sigma_1}_{\Oc(\alpha^2)}
  {}+ \Oc(\alpha^3) \,.
\end{align*}
The two-loop contribution to the form factor is therefore given by
\begin{equation}
\label{eq:F2cont}
  F_2 = F_{\vr,2} + \Sigma_2 + F_{\vr,1} \Sigma_1 \,,
\end{equation}
where $F_{\vr,2}$ is made up of the contributions calculated in
section~\ref{sec:vertex}.
The one-loop corrections are well known:
\begin{align}
\label{eq:Fv1res}
  F_{\vr,1} &= C_F \, \frac{\alpha}{4\pi}
    \left(\frac{\mu^2}{M^2}\right)^\eps S_\eps \,
    \Biggl\{
    \frac{1}{\eps}
    - \lqm^2 + 3 \lqm - \frac{2}{3}\pi^2 - 4
  \nonumber \\* & \qquad
    + \eps \, \biggl[
      \frac{1}{3} \lqm^3 - \frac{3}{2} \lqm^2
      + \left(-\frac{\pi^2}{3} + 8\right) \lqm
      + 2\zeta_3 + \frac{7}{12}\pi^2
  \nonumber \\ & \qquad\quad
      - 12 \biggr]
    + \eps^2 \,\biggl[
      -\frac{1}{12} \lqm^4 + \frac{1}{2} \lqm^3
      + \left(\frac{\pi^2}{12} - 4\right) \lqm^2
  \nonumber \\ & \qquad\quad
      + \left(-4\zeta_3 - \frac{\pi^2}{4} + 16\right) \lqm
      - \frac{13}{180}\pi^4 + \frac{17}{3}\zeta_3 + \pi^2
  \nonumber \\* & \qquad\quad
      - 28
      \biggr]
    \Biggr\}
    + \Oc(\eps^3) + \Oc\left(\frac{M^2}{Q^2}\right)
  ,
\\
\label{eq:Sigma1res}
  \Sigma_1 &= \Sigma_1(p^2=0) =
  C_F \, \frac{\alpha}{4\pi}
    \left(\frac{\mu^2}{M^2}\right)^\eps S_\eps \,
  \Biggl\{ -\frac{1}{\eps} + \frac{1}{2}
  \nonumber \\* & \qquad
    + \eps \left[-\frac{\pi^2}{12} + \frac{1}{4}\right]
    + \eps^2 \left[\frac{1}{3}\zeta_3 + \frac{\pi^2}{24} + \frac{1}{8}\right]
  \Biggr\}
  \nonumber \\* & \quad
  + \Oc(\eps^3)
  \,.
\end{align}
The terms proportional to $\eps$ and $\eps^2$ are needed when
$F_{\vr,1}$ and $\Sigma_1$ are multiplied by other one-loop
contributions containing $1/\eps$-poles from ultraviolet singularities
or $1/\eps^2$-poles from mass singularities.
The sum $F_1 = F_{\vr,1} + \Sigma_1$ constitutes the one-loop form
factor, which is finite at $\eps=0$.

The two-loop self-energy corrections~$\Sigma_2$ originate from the
Feynman diagrams in figures \ref{fig:feynSigmanf},
\ref{fig:feynSigmaAb} and \ref{fig:feynSigmaNA}.%
{\newcommand{\fscale}{0.7}%
\newcommand{\fhspace}{\hspace{1em}}\newcommand{\fvspace}{\\[3ex]}%
\begin{figure}
  \centering
  \includegraphics[scale=\fscale]{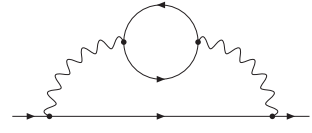}
  \caption{Fermionic self-energy correction}
  \label{fig:feynSigmanf}
\end{figure}%
\begin{figure}
  \centering
  \valignbox[b]{\includegraphics[scale=\fscale]{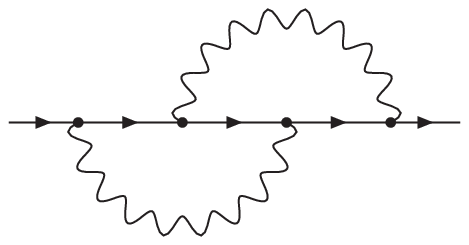}\\(a)}
  \fhspace
  \valignbox[b]{\includegraphics[scale=\fscale]{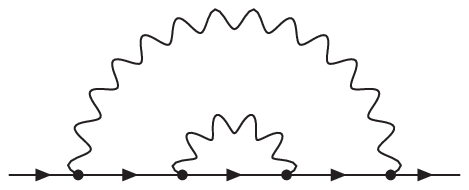}\\(b)}
  \fvspace
  \valignbox[b]{\includegraphics[scale=\fscale]{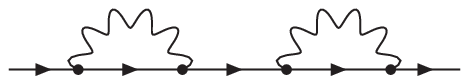}\\(c)}
  \caption{Abelian self-energy corrections}
  \label{fig:feynSigmaAb}
\end{figure}%
\begin{figure}
  \centering
  \valignbox[b]{\includegraphics[scale=\fscale]{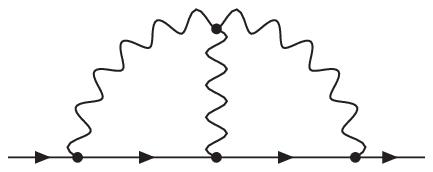}\\(a)}
  \fhspace
  \valignbox[b]{\includegraphics[scale=\fscale]{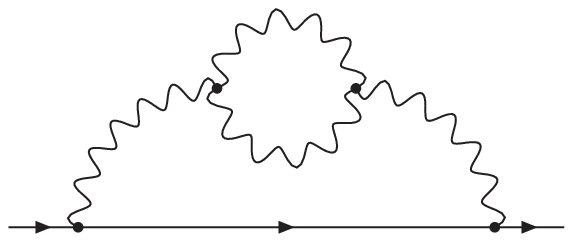}\\(b)}
  \fvspace
  \valignbox[b]{\includegraphics[scale=\fscale]{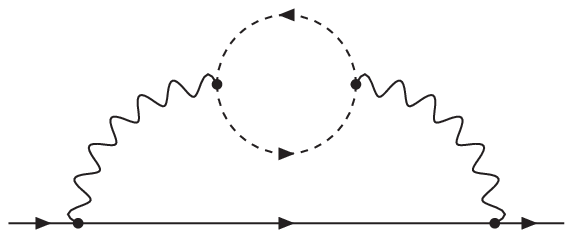}\\(c)}
  \fhspace
  \valignbox[b]{\includegraphics[scale=\fscale]{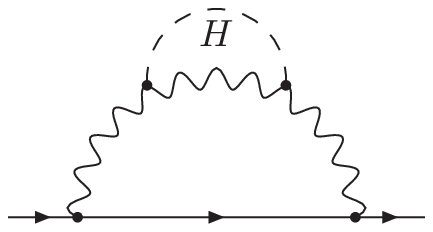}\\(d)}
  \fvspace
  \valignbox[b]{\includegraphics[scale=\fscale]{%
    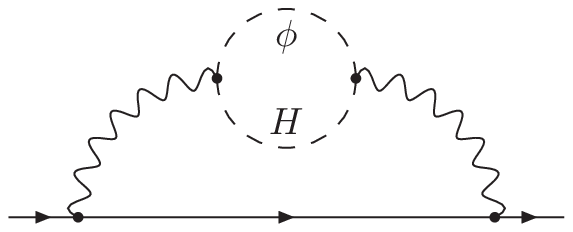}\\(e)}
  \fhspace
  \valignbox[b]{\includegraphics[scale=\fscale]{%
    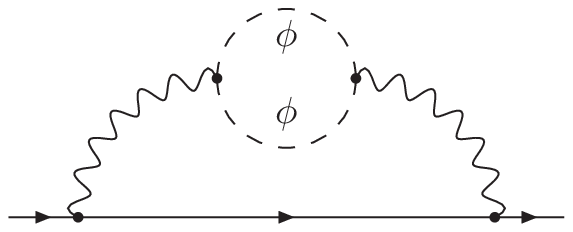}\\(f)}
  \caption{Non-Abelian self-energy corrections}
  \label{fig:feynSigmaNA}
\end{figure}%
}
As for the vertex corrections, ``tadpole'' diagrams are omitted
because their contributions are cancelled by the renormalization of
the gauge boson mass.

The self-energy amplitudes~$\tilde\Sigma$ are quadratic matrices
both in the spinor and in the isospin space.
For massless fermions of momentum~$p$ they are of the form
\begin{equation}
  \tilde\Sigma(p) = -i\dslash p \, \mathbf{1} \, \Sigma(p^2) \,,
\end{equation}
where $\mathbf{1}$ is the unity matrix in the isospin space.
The self-energy correction~$\Sigma$ may be extracted from the
amplitude~$\tilde\Sigma$ by the projection
\begin{equation}
\label{eq:projSigma}
  \Sigma = \frac{i}{4N p^2} \, \Tr(\dslash p \tilde\Sigma) \,,
\end{equation}
where $N=2$ for $SU(2)$ and the trace runs over the spinor and the
isospin indices.
The projection requires $p^2 \ne 0$, whereas we need the self-energies
at $p^2=0$. We have therefore calculated the loop integrals for an
infinitesimally small, but finite~$p^2$. By performing the limit $p^2
\to 0$ before any expansion in~$\eps$, no logarithms~$\ln(p^2)$ appear
and the first two coefficients of a simple Taylor expansion of the
integrals with respect to~$p^2$ are sufficient. The contribution of
every Feynman diagram to the trace in~(\ref{eq:projSigma}) is
proportional to~$p^2$, so no inverse power $1/p^2$ is left.

The reduction of the self-energy diagrams to scalar integrals
using~(\ref{eq:projSigma}) and the calculation of the integrals was
performed similarly to the vertex corrections. In fact, the evaluation
of the self-energy corrections is much simpler because they do not
depend on~$Q^2$, only on $M^2$, and no expansion by regions is needed.
We do not quote further details of this calculation and list only the
total results of each Feynman diagram.

The fermionic self-energy correction in figure~\ref{fig:feynSigmanf}
has already been calculated in~\cite{Feucht:2003yx}:
\begin{align}
\label{eq:Sigmanfres}
  \Sigma_{n_f} &= C_F T_F n_f \left(\frac{\alpha}{4\pi}\right)^2
    \left(\frac{\mu^2}{M^2}\right)^{2\eps} S_\eps^2
    \left( -\frac{1}{\eps} - \frac{1}{2} \right)
  + \Oc(\eps) \,.
\end{align}
The Abelian contributions to the self-energy correction originate from
figure~\ref{fig:feynSigmaAb}a),
\begin{multline}
\label{eq:T1res}
  \Sigma_\Tone =
    \left( C_F^2 - \frac{1}{2} C_F C_A \right)
    \left(\frac{\alpha}{4\pi}\right)^2
    \left(\frac{\mu^2}{M^2}\right)^{2\eps} S_\eps^2
  \\* \times
    \left(
  -\frac{1}{\eps^2}
  + \frac{3}{2\eps}
  - \frac{\pi^2}{2} + \frac{7}{4}
  \right)
  + \Oc(\eps) \,,
\end{multline}
from figure~\ref{fig:feynSigmaAb}b),
\begin{align}
\label{eq:T2res}
  \Sigma_\Ttwo &=
    C_F^2 \left(\frac{\alpha}{4\pi}\right)^2
    \left(\frac{\mu^2}{M^2}\right)^{2\eps} S_\eps^2
    \left(
  \frac{1}{2\eps^2}
  - \frac{1}{4\eps}
  - \frac{\pi^2}{12} + \frac{7}{8}
  \right)
  \nonumber \\* & \quad
  + \Oc(\eps)
  \,,
\end{align}
and from figure~\ref{fig:feynSigmaAb}c), which yields just the square
of the one-loop correction~(\ref{eq:Sigma1res}).
Only the $C_F^2$-part of (\ref{eq:T1res}) belongs to the Abelian
corrections, the $C_F C_A$-part contributes to the non-Abelian corrections.

The other non-Abelian contributions have been evaluated from
figure~\ref{fig:feynSigmaNA}a),
\begin{align}
\label{eq:T1CAres}
  \Sigma_\ToneCA &=
    C_F C_A
    \left(\frac{\alpha}{4\pi}\right)^2
    \left(\frac{\mu^2}{M^2}\right)^{2\eps} S_\eps^2
    \left(
  -\frac{3}{2\eps^2}
  - \frac{5}{4\eps}
  \right)
  \nonumber \\* & \quad
  + \Oc(\eps^0)
  \,,
\end{align}
from the sum of the diagrams with gauge boson and ghost field loops,
figures \ref{fig:feynSigmaNA}b) and \ref{fig:feynSigmaNA}c),
\begin{align}
\label{eq:fcorrWWccres}
  \Sigma_\WWcc &=
    C_F C_A
    \left(\frac{\alpha}{4\pi}\right)^2
    \left(\frac{\mu^2}{M^2}\right)^{2\eps} S_\eps^2
  \cdot \frac{21}{4\eps}
  + \Oc(\eps^0)
  \,,
\end{align}
from figure~\ref{fig:feynSigmaNA}d) with gauge and Higgs boson in the
loop insertion,
\begin{align}
\label{eq:fcorrWHres}
  \Sigma_\WH &=
    \left(\frac{\alpha}{4\pi}\right)^2
    \left(\frac{\mu^2}{M^2}\right)^{2\eps} S_\eps^2
  \left( -\frac{3}{4\eps} \right)
  + \Oc(\eps^0)
  \,,
\end{align}
and, with identical results, from figures \ref{fig:feynSigmaNA}e) and
\ref{fig:feynSigmaNA}f) with Higgs and Goldstone bosons in the loop
insertion,
\begin{align}
\label{eq:fcorrHphiphires}
  \Sigma_\Hphi &= \Sigma_\phiphi =
    \left(\frac{\alpha}{4\pi}\right)^2
    \left(\frac{\mu^2}{M^2}\right)^{2\eps} S_\eps^2
  \cdot \frac{21}{64\eps}
  + \Oc(\eps^0)
  \,.
\end{align}
As for the vertex corrections, the Higgs and Goldstone boson
contributions have been calculated with the approximation $M_H=M$ and
are only valid in a spontaneously broken $SU(2)$ model.
The evaluation of all non-Abelian contributions has been limited to
the poles in~$\eps$ because the non-logarithmic finite term of
order~$\eps^0$ has already been neglected in the calculation of the
corresponding vertex corrections.

\subsection{Coupling constant renormalization}
\label{sec:renalpha}

According to the prescription of the \MSbar{} scheme, the
unrenormalized coupling constant~$\alpha_\bare$ is replaced by the
renormalized coupling~$\alpha$ via
\begin{equation}
\label{eq:alpharen}
  \alpha_\bare =
  \alpha \left( 1 - \frac{\alpha}{4\pi} \, \frac{\beta_0}{\eps} \right)
  + \Oc(\alpha^3) \,,
\end{equation}
where $\beta_0$ is the one-loop coefficient of the renormalization
group $\beta$-function.
$\beta_0$ gets a non-Abelian contribution proportional to~$C_A$, a
fermionic contribution proportional to~$n_f$ and a Higgs
contribution~\cite{Gross:1973id,Politzer:1973fx}:
\begin{equation}
  \beta_0 =
  \frac{11}{3} C_A - \frac{4}{3} T_F n_f - \frac{1}{6} \,.
\end{equation}
As mentioned above, the loop calculations have been performed using
the unrenormalized Feynman rules. Introducing now the renormalized
coupling constant and mass instead of the bare quantities does not
change the two-loop results at order~$\alpha^2$. But the coupling and
mass in the one-loop result have to be regarded as the bare parameters
and must be replaced by the renormalized ones.

By applying the substitution~(\ref{eq:alpharen}) to the one-loop form
factor from equations (\ref{eq:Fv1res}) and (\ref{eq:Sigma1res}), we
get additional contributions of order~$\alpha^2$, namely
\begin{align}
\label{eq:alpharenCAres}
  \Delta F_{C_A}^\alpha &=
    C_F C_A
    \left(\frac{\alpha}{4\pi}\right)^2
    \left(\frac{\mu^2}{M^2}\right)^\eps S_\eps
    \, \Biggl\{
  \nonumber \\* & \qquad
  \frac{1}{\eps} \left[
    \frac{11}{3} \lqm^2 - 11 \lqm + \frac{22}{9}\pi^2 + \frac{77}{6} \right]
  - \frac{11}{9} \lqm^3
  \nonumber \\* & \qquad
  + \frac{11}{2} \lqm^2
  + \left(\frac{11}{9}\pi^2 - \frac{88}{3}\right) \lqm
  - \frac{22}{3}\zeta_3 - \frac{11}{6}\pi^2
  \nonumber \\* & \qquad
  + \frac{517}{12}
  \Biggr\}
  + \Oc(\eps) + \Oc\!\left(\frac{M^2}{Q^2}\right) ,
\\
\label{eq:alpharennfres}
  \Delta F_{n_f}^\alpha &=
    C_F T_F n_f
    \left(\frac{\alpha}{4\pi}\right)^2
    \left(\frac{\mu^2}{M^2}\right)^\eps S_\eps \,
    \Biggl\{
  \nonumber \\* & \qquad
    \frac{1}{\eps} \left[
      -\frac{4}{3} \lqm^2 + 4 \lqm - \frac{8}{9}\pi^2 - \frac{14}{3} \right]
    + \frac{4}{9} \lqm^3 - 2 \lqm^2
  \nonumber \\* & \qquad
      + \left(-\frac{4}{9}\pi^2 + \frac{32}{3}\right) \lqm
      + \frac{8}{3}\zeta_3 + \frac{2}{3}\pi^2 - \frac{47}{3}
    \Biggr\}
  \nonumber \\* & \quad
    + \Oc(\eps) + \Oc\!\left(\frac{M^2}{Q^2}\right) ,
\\
\label{eq:alpharenHiggsres}
  \Delta F_\Higgs^\alpha &=
    \left(\frac{\alpha}{4\pi}\right)^2
    \left(\frac{\mu^2}{M^2}\right)^\eps S_\eps \,
    \Biggl\{
  \nonumber \\* & \qquad
    \frac{1}{\eps} \left[
      -\frac{1}{8} \lqm^2 + \frac{3}{8} \lqm
      - \frac{\pi^2}{12} - \frac{7}{16} \right]
    + \frac{1}{24} \lqm^3
  \nonumber \\* & \qquad
    - \frac{3}{16} \lqm^2
      + \left(-\frac{\pi^2}{24} + 1\right) \lqm
      + \frac{1}{4}\zeta_3 + \frac{\pi^2}{16}
  \nonumber \\* & \qquad
      - \frac{47}{32}
    \Biggr\}
    + \Oc(\eps) + \Oc\!\left(\frac{M^2}{Q^2}\right) .
\end{align}

\subsection{Mass renormalization}
\label{sec:renmass}

The relation between the bare gauge boson mass~$M_\bare$ and the
renormalized mass~$M$ is determined by the gauge boson self-energy
corrections, which have the form
\begin{equation}
  \tilde\Pi^{\mu\nu,ab}(k) =
  i \delta^{ab} \, g^{\mu\nu} k^2 \, \Pi(k^2)
  + \text{terms} \propto k^\mu k^\nu
\end{equation}
at momentum~$k$.
In the on-shell scheme, the square of the physical, renormalized mass
is defined to be the real part of the pole of the propagator.
At one-loop, the relation between $M_\bare$ and $M$ becomes
\begin{equation}
  M_\bare^2 = M^2 \left[ 1 - \Rep \Pi_1(M^2) \right] + \Oc(\alpha^2) \,,
\end{equation}
where $\Pi_1$ is the one-loop contribution to $\Pi$.
This relation leads to the following substitutions in the one-loop
result (\ref{eq:Fv1res}) and (\ref{eq:Sigma1res}):
\begin{align} 
\label{eq:Mreneps}
  \left(\frac{\mu^2}{M^2}\right)^\eps &\to
    \left(\frac{\mu^2}{M^2}\right)^\eps
    \left[ 1 + \eps \, \Rep\Pi_1(M^2) \right]
    + \Oc(\alpha^2) \,,
\\
\label{eq:Mrenlog}
  \lqm^n &\to
    \lqm^n
    + n \, \lqm^{n-1} \, \Rep\Pi_1(M^2)
    + \Oc(\alpha^2) \,,
\end{align}
with $\lqm = \ln(Q^2/M^2)$,
producing additional contributions of order~$\alpha^2$.

The one-loop gauge boson self-energy receives contributions from a
fermion loop,
\begin{multline}
\label{eq:Pinfres}
  \Pi_{n_f}(M^2) =
    T_F n_f \, \frac{\alpha}{4\pi}
    \left(\frac{\mu^2}{M^2}\right)^\eps S_\eps
    \left( -\frac{4}{3\eps} - \frac{20}{9} - \frac{4}{3}i\pi \right)
  \\*
    + \Oc(\eps) \,,
\end{multline}
from the non-Abelian gauge boson and ghost field loops,
\begin{multline}
\label{eq:PiWWccres}
  \Pi_\WWcc(M^2) =
    C_A \, \frac{\alpha}{4\pi} \left(\frac{\mu^2}{M^2}\right)^\eps S_\eps
    \left( \frac{17}{3\eps} - \frac{4\pi}{\sqrt3} + \frac{82}{9} \right)
  \\*
    + \Oc(\eps) \,,
\end{multline}
from the loop with gauge and Higgs boson,
\begin{multline}
\label{eq:PiWHres}
  \Pi_\WH(M^2) =
    \frac{\alpha}{4\pi} \left(\frac{\mu^2}{M^2}\right)^\eps S_\eps
    \left( -\frac{1}{\eps} + \frac{\pi}{\sqrt3} - 2 \right)
  \\*
    + \Oc(\eps) \,,
\end{multline}
and from the loops with Higgs and Goldstone bosons,
\begin{multline}
\label{eq:PiHphiphires}
  \Pi_\Hphi(M^2) = \Pi_\phiphi(M^2) =
  \\*
    \frac{\alpha}{4\pi} \left(\frac{\mu^2}{M^2}\right)^\eps S_\eps
    \left( \frac{5}{12\eps} - \frac{\pi}{4\sqrt3} + \frac{17}{18} \right)
    + \Oc(\eps) \,.
\end{multline}
The self-energy diagrams with ``tadpoles'' have been omitted. They
do not depend on the momentum of the gauge boson, so their
contribution to the mass renormalization cancels exactly the
corresponding vertex correction and field renormalization diagrams
which have already been dropped out before.

Applying the substitutions (\ref{eq:Mreneps}) and (\ref{eq:Mrenlog})
to the one-loop form factor, the self-energy corrections
(\ref{eq:Pinfres})--(\ref{eq:PiHphiphires}) produce the following
contributions to the two-loop form factor:
\begin{align}
\label{eq:Mrennfres}
  \Delta F_{n_f}^M &= C_F T_F n_f \left(\frac{\alpha}{4\pi}\right)^2
    \left(\frac{\mu^2}{M^2}\right)^{2\eps} S_\eps^2 \,
    \Biggl\{ \frac{1}{\eps} \left[ \frac{8}{3} \lqm - 4 \right]
  \nonumber \\* & \qquad
      + \frac{40}{9} \lqm + \frac{4}{3}\pi^2 - \frac{38}{3}
    \Biggr\}
    + \Oc(\eps) + \Oc\!\left(\frac{M^2}{Q^2}\right) ,
\\
\label{eq:MrenWWccres}
  \Delta F_\WWcc^M &=
    C_F C_A
    \left(\frac{\alpha}{4\pi}\right)^2
    \left(\frac{\mu^2}{M^2}\right)^{2\eps} S_\eps^2
    \, \Biggl\{
    \frac{1}{\eps} \left[
      - \frac{34}{3} \lqm + 17 \right]
  \nonumber \\* & \qquad
    + \left(\frac{8\pi}{\sqrt3} - \frac{164}{9}\right) \lqm
    - 4\sqrt{3}\pi - \frac{17}{3}\pi^2
  \nonumber \\* & \qquad
    + \frac{317}{6}
    \Biggr\}
    + \Oc(\eps) + \Oc\!\left(\frac{M^2}{Q^2}\right) ,
\\
\label{eq:MrenWHres}
  \Delta F_\WH^M &=
    \left(\frac{\alpha}{4\pi}\right)^2
    \left(\frac{\mu^2}{M^2}\right)^{2\eps} S_\eps^2
    \, \Biggl\{
  \frac{1}{\eps} \left[
    \frac{3}{2} \lqm - \frac{9}{4} \right]
  \nonumber \\* & \qquad
  + \left( -\frac{1}{2}\sqrt{3}\pi + 3 \right) \lqm
  + \frac{3}{4}\sqrt{3}\pi + \frac{3}{4}\pi^2 - \frac{63}{8}
  \Biggr\}
  \nonumber \\* & \quad
  + \Oc(\eps) + \Oc\!\left(\frac{M^2}{Q^2}\right) ,
\\
\label{eq:MrenHphiphires}
  \Delta F_\Hphi^M &= \Delta F_\phiphi^M =
    \left(\frac{\alpha}{4\pi}\right)^2
    \left(\frac{\mu^2}{M^2}\right)^{2\eps} S_\eps^2
    \, \Biggl\{
  \frac{1}{\eps} \left[
    -\frac{5}{8} \lqm + \frac{15}{16} \right]
  \nonumber \\* & \qquad
  + \left( \frac{1}{8}\sqrt{3}\pi - \frac{17}{12} \right) \lqm
  - \frac{3}{16}\sqrt{3}\pi - \frac{5}{16}\pi^2
  \nonumber \\* & \qquad
  + \frac{113}{32}
  \Biggr\}
  + \Oc(\eps) + \Oc\!\left(\frac{M^2}{Q^2}\right) .
\end{align}

\section{Results and discussion}
\label{sec:result}

The individual results have been presented in the previous sections so
that we can now add them together.
According to the \MSbar{} prescription, the factor
$S_\eps = (4\pi)^\eps e^{-\eps\gamma_E}$ is absorbed into
$\mu^{2\eps}$ by a redefinition of~$\mu$, and we have chosen $\mu=M$ so
that the whole prefactor $(\mu^2/M^2)^\eps S_\eps$ or
$(\mu^2/M^2)^{2\eps} S_\eps^2$ is replaced by 1.
The dependence of the form factor on~$\mu$ can easily be restored by
looking at the running of the coupling~$\alpha$, parametrized
by~$\beta_0$, in the one-loop form factor.
We give the results in $d=4$ dimensions ($\eps=0$) in the
Sudakov limit $Q^2 \gg M^2$.

The fermionic contribution to the Abelian vector form factor is
obtained from equations (\ref{eq:Fvnfres}), (\ref{eq:Sigmanfres}),
(\ref{eq:alpharennfres}) and (\ref{eq:Mrennfres})~\cite{Feucht:2003yx}:
\begin{align}
\label{eq:F2nfres}
  F_{2,n_f} &=
    F_{\vr,n_f} + \Sigma_{n_f} + \Delta F_{n_f}^\alpha + \Delta F_{n_f}^M
  \nonumber \\* &=
  C_F T_F n_f \left(\frac{\alpha}{4\pi}\right)^2 \, \biggl\{
    -\frac{4}{9} \lqm^3
    + \frac{38}{9} \lqm^2
    - \frac{34}{3} \lqm
  \nonumber \\* & \qquad
    + \frac{16}{27}\pi^2 + \frac{115}{9}
    \biggr\} \,,
\end{align}
with $\lqm = \ln(Q^2/M^2)$.
For the Abelian contributions only the $C_F^2$ part of $F_{\vr,\NP}$,
$F_{\vr,\BE}$ and $\Sigma_\Tone$ is considered. The vertex corrections
$F_{\vr,\BE}$ and $F_{\vr,\fc}$ have to be counted twice because two
horizontally mirrored diagrams exist for each of these.
The result follows from equations (\ref{eq:LAres}), (\ref{eq:NPres}),
(\ref{eq:BEres}), (\ref{eq:fcres}), (\ref{eq:Fv1res}),
(\ref{eq:Sigma1res}), (\ref{eq:T1res}) and
(\ref{eq:T2res})~\cite{Feucht:2004rp}:
\begin{align}
\label{eq:F2CFres}
  F_{2,C_F^2} &=
    F_{\vr,\LA} + F_{\vr,\NP}|_{C_F^2} + 2\,F_{\vr,\BE}|_{C_F^2}
    + 2\,F_{\vr,\fc}
  \nonumber \\* & \qquad
    + \Sigma_\Tone|_{C_F^2} + \Sigma_\Ttwo + (\Sigma_1)^2
    + F_{\vr,1} \Sigma_1
  \nonumber \\ &=
  C_F^2
  \left(\frac{\alpha}{4\pi}\right)^2
  \, \biggl\{
  \frac{1}{2} \lqm^4
  - 3 \lqm^3
  + \left(\frac{2}{3}\pi^2 + 8\right) \lqm^2
  \nonumber \\* & \qquad
  - \Bigl(-24\zeta_3 + 4\pi^2 + 9\Bigr) \, \lqm
  + 256\,\Li4\!\left(\frac{1}{2}\right)
  \nonumber \\ & \qquad
  + \frac{32}{3}\ln^4{2} - \frac{32}{3}\pi^2\ln^2{2}
  - \frac{52}{15}\pi^4 + 80\zeta_3
  \nonumber \\* & \qquad
  + \frac{52}{3}\pi^2
  + \frac{25}{2}
  \biggr\}
  \,.
\end{align}
For the non-Abelian contributions proportional to~$C_F C_A$ the
remaining part of $F_{\vr,\NP}$, $F_{\vr,\BE}$ and $\Sigma_\Tone$ is
considered together with the purely non-Abelian results from equations
(\ref{eq:BECAres}), (\ref{eq:WWccres}), (\ref{eq:T1CAres}),
(\ref{eq:fcorrWWccres}), (\ref{eq:alpharenCAres}) and
(\ref{eq:MrenWWccres}):
\begin{align}
\label{eq:F2CFCAres}
  F_{2,C_F C_A} &=
    \Bigl[ F_{\vr,\NP} + 2\,F_{\vr,\BE} + \Sigma_\Tone
      \Bigr]_{C_F C_A}
  \nonumber \\* & \qquad
    + 2\,F_{\vr,\BECA} + F_{\vr,\WWcc}
    + \Sigma_\ToneCA + \Sigma_\WWcc
  \nonumber \\* & \qquad
    + \Delta F_{C_A}^\alpha
    + \Delta F_\WWcc^M
  \nonumber \\ &=
    C_F C_A
    \left(\frac{\alpha}{4\pi}\right)^2
    \, \biggl\{
  \frac{11}{9} \lqm^3
  - \left(-\frac{\pi^2}{3} + \frac{233}{18}\right) \lqm^2
  \nonumber \\* & \qquad
  + \biggl( 4\sqrt{3}\,\Cl2\!\left(\frac{\pi}{3}\right)
    + \frac{8\pi}{\sqrt3} - \frac{88}{3}\zeta_3 + \frac{11}{9}\pi^2
  \nonumber \\* & \qquad\qquad
    + \frac{193}{6} \biggr) \, \lqm
  \biggr\}
  + \Oc(\lqm^0)
  \,.
\end{align}
The Higgs contribution results from equations (\ref{eq:WHvertexres}),
(\ref{eq:Hphiphivertexres}), (\ref{eq:fcorrWHres}),
(\ref{eq:fcorrHphiphires}), (\ref{eq:alpharenHiggsres}),
(\ref{eq:MrenWHres}) and (\ref{eq:MrenHphiphires}):
\begin{align}
\label{eq:F2Higgsres}
  F_{2,\Higgs} &=
    F_{\vr,\WH} + F_{\vr,\Hphi} + F_{\vr,\phiphi}
    + \Sigma_\WH + \Sigma_\Hphi + \Sigma_\phiphi
  \nonumber \\* & \qquad
    + \Delta F_\Higgs^\alpha
    + \Delta F_\WH^M + \Delta F_\Hphi^M + \Delta F_\phiphi^M
  \nonumber \\ &=
    \left(\frac{\alpha}{4\pi}\right)^2 \,
    \biggl\{
    -\frac{1}{24} \lqm^3 + \frac{25}{48} \lqm^2
    - \biggl( -\frac{1}{2}\sqrt{3}\,\Cl2\!\left(\frac{\pi}{3}\right)
  \nonumber \\* & \qquad
    + \frac{1}{4}\sqrt{3}\pi + \frac{\pi^2}{24}
    + \frac{23}{16} \biggr) \, \lqm
    \biggr\}
    + \Oc(\lqm^0)
  \,.      
\end{align}
The two non-Abelian contributions $F_{2,C_F C_A}$ and $F_{2,\Higgs}$
depend on the Feynman--'t~Hooft gauge in which they have been
calculated. Only their sum is gauge invariant:
\begin{align}
\label{eq:F2NAres}
  &\!\!\! F_{2,\NA} = F_{2,C_F C_A} + F_{2,\Higgs}
  \nonumber \\* &=
    \left(\frac{\alpha}{4\pi}\right)^2 \,
    \biggl\{
    \frac{43}{24} \lqm^3
    - \left( -\frac{\pi^2}{2} + \frac{907}{48} \right) \lqm^2
  \nonumber \\ & \qquad
    + \biggl( \frac{13}{2}\sqrt{3}\,\Cl2\!\left(\frac{\pi}{3}\right)
      + \frac{15}{4}\sqrt{3}\pi
      - 44\zeta_3
      + \frac{43}{24}\pi^2
  \nonumber \\* & \qquad\qquad
      + \frac{749}{16} \biggr) \, \lqm
    \biggr\}
    + \Oc(\lqm^0)
  \,,
\end{align}
where the values $C_F = 3/4$ and $C_A = 2$ for the $SU(2)$ gauge group
have been used.
Adding all contributions together, the two-loop form factor is given
by~\cite{Jantzen:2005xi}
\begin{align}
\label{eq:F2res}
  F_2 &= F_{2,n_f} + F_{2,C_F^2} + F_{2,\NA}
  \nonumber \\* &=
  \left(\frac{\alpha}{4\pi}\right)^2 \biggl\{
    \frac{9}{32} \lqm^4
    + \left(\frac{5}{48} - \frac{n_f}{6}\right) \lqm^3
  \nonumber \\* & \qquad
    + \left(\frac{7}{8}\pi^2 - \frac{691}{48} + \frac{19}{12} n_f\right)
      \lqm^2
    + \biggl( \frac{13}{2}\sqrt{3}\,\Cl2\!\left(\frac{\pi}{3}\right)
  \nonumber \\* & \qquad\quad
      + \frac{15}{4}\sqrt{3}\pi
      - \frac{61}{2}\zeta_3
      - \frac{11}{24}\pi^2
      + \frac{167}{4}
      - \frac{17}{4} n_f \biggr) \, \lqm
    \biggr\}
  \nonumber \\* & \qquad
    + \Oc(\lqm^0)
  \,.
\end{align}
The coefficients of the first three logarithms $\lqm^4$, $\lqm^3$ and
$\lqm^2$ agree with the NNLL prediction of the evolution equation
approach~\cite{Kuhn:2001hz,Kuhn:2001hzE}.
The coefficient of the linear logarithm is a new result.

Let us have a look at the numerical size of the coefficients in the
individual contributions. For the fermionic contribution we set
$n_f=6$ for 3~lepton and 3$\times$3 quark doublets from which only the
left-handed degrees of freedom couple to the gauge bosons.
\begin{alignat}{10}
\label{eq:F2numlogs}
  F_{2,n_f} &\approx \left(\frac{\alpha}{4\pi}\right)^2 \bigl( &
    &&{} - 1.0\,&\lqm^3 &{} + 9.5\,&\lqm^2 &{} - 26\,&\lqm
    + 42 \bigr) \,, \nonumber \\*
  F_{2,C_F^2} &\approx \left(\frac{\alpha}{4\pi}\right)^2 \bigl( &
    0.3\,&\lqm^4 &{} - 1.7\,&\lqm^3 &{} + 8.2\,&\lqm^2 &{} - 11\,&\lqm
    + 15 \bigr) \,, \nonumber \\*
  \lefteqn{\hspace*{-2.5em} F_{2,\NA}} \nonumber \\*
    &\approx \left(\frac{\alpha}{4\pi}\right)^2 \bigl( &
    &&{} 1.8\,&\lqm^3 &{} - 14\,&\lqm^2 &{} + 43\,&\lqm
    + \ldots \bigr)
\end{alignat}
We notice that all three contributions show a similar pattern of
coefficients with alternating signs and growing size.
At a typical energy in the TeV range, $Q=1\,$TeV, using $M=80\,$GeV
and $\alpha/(4\pi)=0.003$ as rough values for the weak interaction,
the individual logarithmic terms have the following numerical size in
per mil (1/1000): 
\begin{alignat}{12}
\label{eq:F2numcont}
  && \lqm^4 && \lqm^3 && \lqm^2 && \lqm^1 && \lqm^0
    \nonumber \\*
  F_{2,n_f} &\to& &&{} -1.2 &&{} +2.2 &&{} -1.2 &&{} +0.4
    & \,, \nonumber \\*
  F_{2,C_F^2} &\to&{} +1.6 &&{} -2.0 &&{} +1.9 &&{} -0.5 &&{} +0.1
    & \,, \nonumber \\*
  F_{2,\NA} &\to& &&{} +2.1 &&{} -3.2 &&{} +2.0 &&{} + \ldots
\end{alignat}
The pattern of growing coefficients with alternating signs produces
large cancellations between the terms of different powers of
logarithms and also between $F_{2,n_f}$, $F_{2,C_F^2}$ and $F_{2,\NA}$.
In each line of~(\ref{eq:F2numcont}), the largest term is reached at
the quadratic or (for $F_{2,C_F^2}$) already at the cubic logarithm.
The linear-logarithmic term is less significant and, at least for
the fermionic and the Abelian part, the non-logarithmic constant is
again smaller by a factor of 3 or more.
For the sum of the three contributions,
\begin{alignat}{10}
\label{eq:F2totnumcont}
  && \lqm^4 && \lqm^3 && \lqm^2 && \lqm^1
    \nonumber \\*
  F_2 &\to&{} +1.6 &&{} -1.0 &&{} +0.9 &&{} +0.3 & \,,
\end{alignat}
the logarithmic terms are monotonically decreasing in size already
from $\lqm^4$ on.
Due to the cancellations between the individual contributions
in~(\ref{eq:F2numcont}), the non-logarithmic constant of $F_{2,n_f}$
is larger than the total linear-logarithmic term
in~(\ref{eq:F2totnumcont}). But the logarithmic terms in all
contributions and in the total form factor are getting significantly
smaller from the linear logarithm on. So we do not expect the
neglected non-logarithmic constant of the total result to be larger
than the total linear-logarithmic term.
This leads us to the conclusion that the N$^3$LL result with all
logarithmic terms approximates well the full result.

\begin{figure}
  \centering
  \includegraphics[width=\columnwidth]{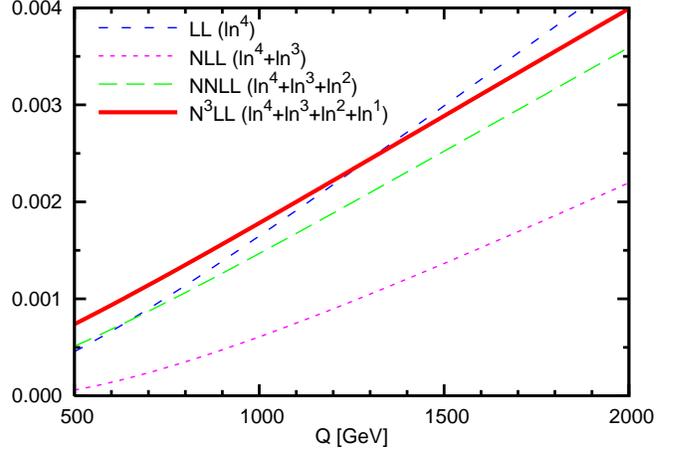}
  \caption{Two-loop contribution to the Abelian vector form
    factor~$F_2$ in successive logarithmic approximations,
    using the values $M=80\,$GeV and $\alpha/(4\pi)=0.003$}
  \label{fig:F2logs}
\end{figure}

Figure~\ref{fig:F2logs} illustrates the behaviour of the successive
logarithmic approximations, starting from the LL approximation with
only the $\lqm^4$ term and adding one after the other the smaller
powers of logarithms.%

The result presented here relies on the approximation that the Higgs
mass is equal to the gauge boson mass, $M_H=M$.
In order to investigate the dependence of the form factor on the Higgs
mass, we have also calculated the Higgs contributions in the
hypothetical case of a vanishing Higgs mass, $M_H=0$.
Then equation~(\ref{eq:F2Higgsres}) becomes
\begin{align}
\label{eq:F2HiggsMH0res}
  F_{2,\Higgs}^{M_H=0} &=
    \left(\frac{\alpha}{4\pi}\right)^2 \,
    \biggl\{
    -\frac{1}{24} \lqm^3 + \frac{25}{48} \lqm^2
    - \biggl( \frac{3}{4}\sqrt{3}\,\Cl2\!\left(\frac{\pi}{3}\right)
  \nonumber \\* & \qquad
      - \frac{1}{8}\sqrt{3}\pi -  \frac{3}{16}\pi^2
      + \frac{25}{16} \biggr) \, \lqm
    \biggr\}
    + \Oc(\lqm^0)
  \,.      
\end{align}
Only the coefficient of the linear logarithm differs between equations
(\ref{eq:F2Higgsres}) and (\ref{eq:F2HiggsMH0res}). The coefficients
of the cubic and quadratic logarithms are the same, they do not depend
on the Higgs boson mass and have already been determined in the
evolution equation approach~\cite{Kuhn:2001hz,Kuhn:2001hzE}.
By setting $M_H=0$, the coefficient of the linear logarithm of
$F_{2,\NA}$ in~(\ref{eq:F2numlogs}) numerically changes from 43 to 45,
and the contribution of this term in~(\ref{eq:F2numcont}) is shifted
from $2.0$ to $2.1$.
So the variation of the N$^3$LL form factor between the two cases
$M_H=M$ and $M_H=0$ is smaller than the total linear-logarithmic
contribution by a factor of~3.
On this basis we expect the deviation of the form factor with the true
Higgs mass from our result to be comparable to the neglected
non-logarithmic constant.

Altogether we estimate the accuracy of our form factor result to be of
the order of the linear-logarithmic contribution, i.e. about half a
per mil with respect to the Born result.

The result for the Abelian vector form factor presented
in~(\ref{eq:F2res}) has been combined
in~\cite{Jantzen:2005xi,Jantzen:2005az} with the reduced amplitude
from equation~(\ref{eq:Adecomp}) in order to obtain the four-fermion
scattering amplitude in the spontaneously broken $SU(2)$ model in
N$^3$LL accuracy.
In addition predictions for the electroweak model have been obtained
by separating the infrared-divergent electromagnetic contributions
(cf. appendix~\ref{sec:massgap}) and by expanding in the mass
difference between the $W$ and $Z$ bosons.
For a discussion of this procedure and of the accuracy of the
electroweak corrections we refer
to~\cite{Jantzen:2005xi,Jantzen:2005az}.

\section{Summary}
\label{sec:summary}

In the present paper we have discussed in detail the calculation of
the two-loop corrections to the Abelian vector form factor in a
spontaneously broken $SU(2)$ model.
The result was obtained in N$^3$LL accuracy and contains all
logarithmically enhanced terms. It enables the derivation of
electroweak corrections to four-fermion processes with an error of a
few per mil to one percent, thus coping with the expected experimental
accuracy at a future linear collider.

\begin{acknowledgement}
  \emph{Acknowledgements.}
  We would like to thank Johann H. K\"uhn and Alexander A. Penin for the
  fruitful collaboration in the N$^3$LL calculation of the
  four-fermion processes and for reading the manuscript.
  The work of B.J. was supported in part by Cusanuswerk,
  Lan\-des\-gra\-duier\-ten\-f\"or\-derung Baden-W\"urttemberg and the
  DFG Gra\-duier\-ten\-kol\-leg ``Hochenergiephysik und
  Teil\-chen\-astro\-physik''.
  The work of V.A.S. was supported in part by the Russian Foundation
  for Basic Research through project \mbox{05-02-17645} and DFG
  Mercator Grant No.~\mbox{Ha 202/110-1}.
  The work of both authors was supported by the
  Sonderforschungsbereich Transregio~9.
\end{acknowledgement}

\appendix

\section{Feynman rules}
\label{sec:feynman}

This appendix lists the Feynman rules of the vertices which are needed
for the calculation of the form factor, as they follow from the
Lagrangian of the spontaneously broken $SU(2)$ gauge model described
in section~\ref{sec:formfactor}.

The gauge boson fields of mass~$M=M_W$ are $W^a_\mu$, $a=1,2,3$ (with
Lorentz vector index $\mu$). To each $W^a$ corresponds a ghost
field~$c^a$ (and antighost~$\bar c^a$) and a Goldstone boson~$\phi^a$,
one of the unphysical components of the Higgs doublet.
In the Feynman--'t~Hooft gauge used by us, there is $M_c = M_\phi = M_W$.
The physical Higgs boson~$H$ has the mass~$M_H$.
Finally, $\psi$ denotes a fermion (lepton or quark) doublet of Dirac
spinors, and $g$ is the weak $SU(2)$ coupling.

Vertices involving four fields and vertices without a gauge boson
do not appear in our present calculation and are omitted here.

\paragraph{Gauge boson coupling to fermions}
\[
  \vcentergraphics{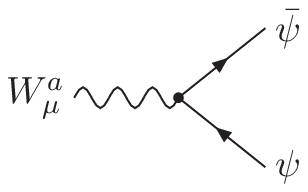} = ig \gamma_\mu t^a
\]

\paragraph{Gauge boson self-coupling}
\[
  \vcentergraphics{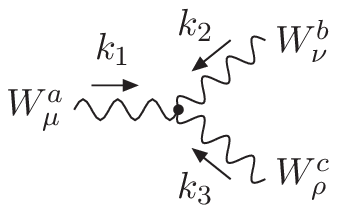} =
  \begin{cases}
    \begin{aligned}
      g\,f^{abc} \Bigl[ \, &
        g_{\mu\nu} (k_1-k_2)_\rho
      \\
        + \, & g_{\nu\rho} (k_2-k_3)_\mu
      \\
        + \, & g_{\rho\mu} (k_3-k_1)_\nu
      \Bigr]
    \end{aligned}
  \end{cases}
\]

\paragraph{Gauge boson coupling to ghost fields}
\[
  \vcentergraphics{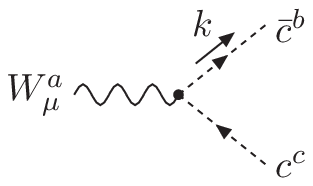} = -g\,f^{abc} k_\mu
\]

\paragraph{Gauge boson coupling to Higgs and Goldstone bosons}
\begin{align*}
  \vcentergraphics{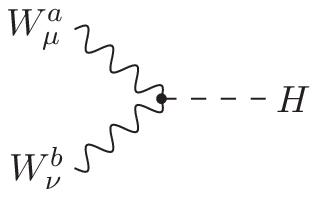} &= ig M\,g_{\mu\nu} \delta^{ab}
\\[1ex]
  \vcentergraphics{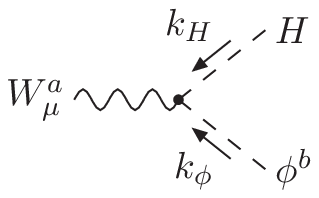} &= \frac{g}{2}
    \delta^{ab} (k_H-k_\phi)_\mu
\\[1ex]
  \vcentergraphics{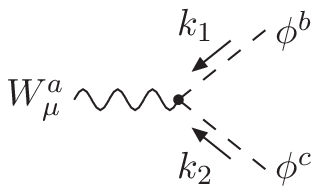} &= -\frac{g}{2}
    f^{abc} (k_1-k_2)_\mu
\end{align*}
In contrast to the other vertices above, which can be used in any
$SU(N)$ gauge model, the couplings involving Higgs and Goldstone
bosons are only valid for the spontaneously broken $SU(2)$ model.

\section{Expansion by regions}
\label{sec:regions}

The asymptotic expansion of Feynman integrals in limits
typical of Euclidean space is given by well-known
prescriptions as a sum over certain subgraphs~\cite{%
  Chetyrkin:1988zz,Chetyrkin:1988cu,Gorishnii:1989dd,%
  Smirnov:1990rz,Smirnov:1994tg}.
However, the Sudakov limit we are dealing with
is typical of Minkowski space. Still for some special
cases, similar graph-theoretical prescriptions were
obtained~\cite{Smirnov:1996ng,Czarnecki:1996nr,Smirnov:1997gx}.
In particular, as it was shown in \cite{Smirnov:1997gx}, they can be
applied to expand the planar two-loop diagram of
figure~\ref{fig:feynAb}a) in the Sudakov limit.

The bad news is that for other relevant diagrams, such as the
non-planar and the Mercedes--Benz diagrams, graph-theoretical
prescriptions are not available.
The good news is that one can apply here (and for any other limit) the
strategy of expansion by regions~\cite{%
  Beneke:1997zp,Smirnov:1998vk,Smirnov:1999bz,Smirnov:2002pj}
which consists of the following prescriptions:
\begin{itemize}
\item
  Divide the space of the loop momenta into various regions and, in
  every region, expand the integrand in a Taylor series with respect
  to the parameters that are considered small there.
\item
  Integrate the integrand, expanded in the appropriate way in every
  region, over the \emph{whole integration domain} of the loop momenta.
\item
  Set to zero any scaleless integral.
\end{itemize}
To apply this strategy to a given limit one should first understand,
using various examples, which regions are relevant to it.
In the Sudakov limit under consideration, these are the following
regions, for a loop momentum~$k$:
\begin{align*}
  \text{\emph{hard} (h):} \quad & k \sim Q \,, \\*
  \text{\emph{1-collinear} (1c):} \quad & k_+ \sim \frac{M^2}{Q} \,,\quad
    k_- \sim Q \,,\quad k_\bot \sim M \,, \\*
  \text{\emph{2-collinear} (2c):} \quad & k_+ \sim Q \,,\quad
    k_- \sim \frac{M^2}{Q} \,,\quad k_\bot \sim M \,, \\
  \text{\emph{soft} (s):} \quad & k \sim M \,, \\*
  \text{\emph{ultrasoft} (us):} \quad & k \sim \frac{M^2}{Q} \,.
\end{align*}
By $k\sim Q$ etc. we mean that any component of the vector~$k$ is of
order~$Q$, and $k_\pm$, $k_\bot$ are the components of~$k$ defined
after equation~(\ref{eq:def2c}).
In other versions of the Sudakov limit, ultracollinear regions can also
participate~\cite{Smirnov:1999bz}, but they are irrelevant to the
present version.

So we obtain with this strategy the asymptotic expansion of our
integrals as a sum of contributions generated by various regions. For
brevity, we omit the word ``generated'' and speak about contributions
of regions, although the integration in each contribution is performed
over the whole space of the loop momenta.

In fact, this strategy is a generalization of the original strategy
based on similar regions~\cite{%
  Sterman:1986aj,Mueller:1979ih,Collins:1989bt},
where the cut-offs specifying the regions were not removed so that the
integrations were bounded by the regions under consideration.

\section{Mellin--Barnes representation}
\label{sec:MB}

The Mellin--Barnes representation is a powerful tool for solving
two closely related problems:
a)~The calculation of Feynman integrals.
b)~The asymptotic expansion of Feynman integrals in various
kinematical limits.

The basic identity of the Mellin--Barnes representation is the
following (valid for $|\arg X - \arg Y| < \pi$):
\begin{equation}
\label{eq:MB}
  \frac{1}{(X+Y)^\lambda} =
  \frac{1}{\Gamma(\lambda)}
  \MBint z \, \Gamma(-z) \Gamma(\lambda+z) \,
  \frac{Y^z}{X^{\lambda+z}} \,.
\end{equation}
It replaces a sum raised to any power by individual factors which are
raised to powers depending on the Mellin--Barnes parameter~$z$. This
simplification of the structure is obtained at the cost of an
additional integration.
The integration contour in the Mellin--Barnes integrals runs from
$-i\infty$ to $+i\infty$ and is chosen in such a way that poles from
gamma functions of the form $\Gamma(\ldots+z)$ lie on the left hand
side of the contour (``left poles'') and poles from gamma functions
of the form $\Gamma(\ldots-z)$ lie on the right hand side of the
contour (``right poles'').
The contour cannot always be chosen as a straight line, especially if
$\Rep\lambda < 0$.

If $|X|<|Y|$, the integration contour can be closed on the left hand
side at $\Rep z = -\infty$, and the integral is given by the sum over
the residues at the left poles. This sum corresponds to the Taylor
expansion of $1/(X+Y)^\lambda$ for $|X|<|Y|$.
On the other hand, if $|X|>|Y|$, the contour can be closed on the right
hand side at $\Rep z = +\infty$, and the sum over the residues at the
right poles corresponds to the Taylor expansion for $|X|>|Y|$.
In the limiting case $|X|=|Y|$ the Mellin--Barnes integral is also
convergent and is given by the sum of the residues on either the left
or the right side of the contour -- provided that these sums converge,
which is often the case, especially when more than two gamma functions
are present.

The first application of the Mellin--Barnes representation was,
probably, in \cite{Usyukina:1975yg}.
The simplest possibility of using it is to transform massive
propagators ($X=k^2$, $Y=-M^2$) into massless ones (see e.g.
\cite{Boos:1990rg,Davydychev:1990jt,Davydychev:1990cq} as early
references).
In general, one starts from Feynman, alpha or Schwinger parameters and
uses the Mellin--Barnes representation to separate arbitrary terms
raised to some powers in such a way that the resulting parametric
integrals can be calculated in terms of gamma functions (see e.g.
\cite{Greub:1996tg,Greub:2000sy,Asatryan:2001zw,Bieri:2003ue}).
In the context of dimensional regularization, when the explicit evaluation
at general values of $d=4-2\eps$ is hardly possible and
one is oriented at calculating Feynman integrals in a Laurent expansion
in $\eps$, the systematic evaluation by Mellin--Barnes representations
was initiated in \cite{Smirnov:1999gc,Tausk:1999vh}.
An essential step of the evaluation procedure is the resolution of
singularities in $\eps$, with the goal to represent a given multiple
Mellin--Barnes integral as a sum of integrals where
the Laurent expansion of the integrands becomes possible.
This is achieved by taking residues and shifting contours.
Two different strategies for implementing this step were suggested
in \cite{Smirnov:1999gc} and \cite{Tausk:1999vh},
respectively.

The identity~(\ref{eq:MB}) is valid for all powers~$\lambda$. In fact,
the crucial point is not the convergence of the integral in the basic
identity~(\ref{eq:MB}), but the interchange of the order of
integrations between the Mellin--Barnes integral and the (Feynman,
alpha or Schwinger) parameter integrals. The necessary convergence of
the parameter integrals restricts the real part of the Mellin--Barnes
parameter~$z$ to a specific range. If this range has a non-empty
overlap with the interval $(-\Rep\lambda,0)$, the integration
contour over~$z$ can be chosen as a straight line parallel to the
imaginary axis within the allowed range on the real axis. One can
find values for the power~$\lambda$ and other parameters such that
an allowed range for the real part of~$z$ exists. The analytic
continuation to the desired parameter values is then obtained by
accounting for the residues which cross the fixed integration contour
when the parameter values are smoothly changed~\cite{Tausk:1999vh}.
Alternatively, the contour of the Mellin--Barnes integration can be
deformed in such a way that it separates the poles of gamma functions
with a ``$+z$'' dependence from the ones with a ``$-z$'' dependence
even for the desired parameter values~\cite{Smirnov:1999gc}. Not only
the gamma functions from the Mellin--Barnes
representation~(\ref{eq:MB}), but also the ones introduced by the
evaluation of the parameter integrals have to be considered here.
As long as the prescription following equation~(\ref{eq:MB}) for the
integration contour is respected, the convergence of the integrals is
provided and all residues are accounted for on the correct side of the
contour.
This is still true if multiple Mellin--Barnes integrals are introduced
by the iterated application of~(\ref{eq:MB}).
Even when $\lambda$ is a non-positive integer and $\Gamma(\lambda)$ in
the denominator gets singular, the right hand side of~(\ref{eq:MB}) is
given by the limit where $\lambda$ approaches its actual value. In
this case only a finite number of residues give non-vanishing
contributions and reproduce the Binomial formula for
$(X+Y)^{|\lambda|}$.

Often the Mellin--Barnes representation is used for asymptotic
expansions (see e.g.~\cite{Greub:1996tg,Greub:2000sy,Asatryan:2001zw,%
Bieri:2003ue,Smirnov:2002mg,Friot:2005cu}). When the Mellin--Barnes
integral contains the factor~$t^z$ with some parameter~$t$, the
asymptotic expansion in the limit $t \to 0$ is given by picking up the
residues on the right hand side of the integration contour. The
asymptotic expansion in the limit $t \to \infty$ is given by the
residues on the left hand side of the contour.
By expanding $t^z$ in $z$ about the poles of the integrand, the
explicit form of the asymptotic expansion in powers of $t$ and $\ln t$
with coefficients from the Laurent expansion of the Mellin--Barnes
integrand can easily be obtained~\cite{Friot:2005cu}.
In practice, however, the most adequate way to perform the asymptotic
expansion depends on the specific problem. For the work presented in
this paper we have applied the method of expansion by regions
(appendix~\ref{sec:regions}) and used Mellin--Barnes representations in
the purpose of asymptotic expansion as a cross-check. When calculating
scalar integrals for general propagator powers~$n_i$ as in
section~\ref{sec:vertex}, the leading contributions can be obtained
from the Mellin--Barnes representation by taking the residue at the
first pole of each gamma function on the correct side of the
integration contour. It turned out that in many cases the expressions
obtained by the expansion of the loop integral within the expansion by
regions method were simpler than the expressions extracted from the
Mellin--Barnes representation of the full integral.

If the Mellin--Barnes representation is applied to the calculation of
Feynman integrals (in particular, of individual contributions in an
asymptotic expansion, as in the present work), when no large or small
parameter~$t$ is present as $t^z$ in the Mellin--Barnes integrals or
when the full dependence on~$t$ is desired, all residues on one side
of the integration contour have to be considered and summed up.
Some integrations in multiple Mellin--Barnes integrals can be performed
explicitly by the application of identities based on the first Barnes
lemma~\cite{Barnes:1908},
\begin{multline}
\label{eq:Barnes1}
  \MBint z \,
  \Gamma(\lambda_1+z) \Gamma(\lambda_2+z)
  \Gamma(\lambda_3-z) \Gamma(\lambda_4-z)
\\* =
  \frac{\Gamma(\lambda_1+\lambda_3) \Gamma(\lambda_1+\lambda_4)
    \Gamma(\lambda_2+\lambda_3) \Gamma(\lambda_2+\lambda_4)}
    {\Gamma(\lambda_1+\lambda_2+\lambda_3+\lambda_4)}
  \,,
\end{multline}
or on the second Barnes lemma~\cite{Barnes:1910},
\begin{multline}
\label{eq:Barnes2}
  \MBint z \,
  \frac{\Gamma(\lambda_1+z) \Gamma(\lambda_2+z) \Gamma(\lambda_3+z)}
    {\Gamma(\lambda_1+\lambda_2+\lambda_3+\lambda_4+\lambda_5+z)}
  \\* \shoveright{\times
  \Gamma(\lambda_4-z) \Gamma(\lambda_5-z)}
  \\ \shoveleft{=
  \frac{\Gamma(\lambda_1+\lambda_4) \Gamma(\lambda_1+\lambda_5)
    \Gamma(\lambda_2+\lambda_4) \Gamma(\lambda_2+\lambda_5)}
    {\Gamma(\lambda_1+\lambda_2+\lambda_4+\lambda_5)
    \Gamma(\lambda_1+\lambda_3+\lambda_4+\lambda_5)}}
  \\* \times
  \frac{\Gamma(\lambda_3+\lambda_4) \Gamma(\lambda_3+\lambda_5)}
    {\Gamma(\lambda_2+\lambda_3+\lambda_4+\lambda_5)}
\end{multline}
(see a collection of such formulae in Appendix~D of
\cite{Smirnov:2004ym}).

Mellin--Barnes integrals develop singularities when a left pole and a
right pole glue together in one point for some limit, e.g. $\eps\to0$
from dimensional regularization. These singularitities are directly
present in the formulae (\ref{eq:Barnes1}) and (\ref{eq:Barnes2}) of
the first and second Barnes lemma.
In more complicated cases, it is usually a good idea to first extract
the potentially singular residues by shifting the integration
contours~\cite{Smirnov:1999gc} or by an analytic continuation as
described above and in~\cite{Tausk:1999vh}. Then the integrand may be
expanded in the desired limits of its parameters. For an analytical
result the residues on one side of the integration contour are summed
up with the help of computer algebra programs, summation tables (see
e.g.~\cite{Smirnov:2004ym}) or algorithms
like~\cite{Moch:2001zr,Weinzierl:2004bn}.

Characteristic examples of recent sophisticated calculations based on
the technique of Mellin--Barnes representations can be found in
\cite{Smirnov:2003vi,Bern:2005iz}.
These results were crucial to check (in~\cite{Bern:2005iz}) cross
order relations in $N=4$ supersymmetric Yang--Mills theory conjectured
in~\cite{Anastasiou:2003kj}.
Very recent results on checking the iteration structure in this theory
with the help of Mellin--Barnes representations have been obtained
in~\cite{Cachazo:2006mq,Cachazo:2006tj,Bern:2006vw}.

Also recently algorithms for the automatic evaluation of Mellin--Barnes
integrals have been formulated~\cite{Anastasiou:2005cb,Czakon:2005rk}.
These rely on the strategy of~\cite{Tausk:1999vh} for the analytic
continuation in the parameter~$\eps$. The algorithms provide a basis
for the analytic evaluation, and at least they can be applied, in
their present form, to the numerical evaluation.
The algorithm of~\cite{Czakon:2005rk} is already implemented in
\mbox{\textsc{Mathematica}} and would have been applied by us at least for
numerical checks if it had existed early enough.

\section{Contributions in a theory with a mass gap}
\label{sec:massgap}

The separation of the infrared-divergent electromagnetic contributions
as described in~\cite{Feucht:2004rp} requires the two-loop corrections
in the combined $SU(2)\times U(1)$ or $U(1) \times U(1)$ theory with
massive and massless gauge bosons.
In addition to the results presented in the sections \ref{sec:vertex}
and \ref{sec:ren} of this paper, two-loop vertex and self-energy
corrections with one massive $SU(2)$ or $U(1)$ gauge boson and one
massless $U(1)$ gauge boson are needed.

As we regard an $SU(2)\times U(1)$ model without mixing between the
two gauge groups (see \cite{Jantzen:2005xi,Jantzen:2005az} for a
discussion of this aspect), only the Abelian vertex and self-energy
diagrams (figures \ref{fig:feynAb} and \ref{fig:feynSigmaAb})
contribute. After replacing one of the two massive $SU(2)$ gauge bosons
in these diagrams by a massless $U(1)$ gauge boson, we obtain the
results listed in the following paragraphs.

The \emph{planar vertex correction} of the diagram in
figure~\ref{fig:feynAb}a) with line~5 (cf. figure~\ref{fig:scalarLA})
massless is
\begin{align}
\label{eq:LAm6res}
  F_{\vr,\LA}^{M_5=0} &=
    C_F
    \, \frac{\alpha \alpha'}{(4\pi)^2}
    \left(\frac{\mu^2}{M^2}\right)^{2\eps} S_\eps^2
    \, \Biggl\{
  -\frac{2}{\eps^3}
  \nonumber \\* & \quad
  + \frac{1}{\eps^2} \left[
    2 \lqm^2 - 4 \lqm + \frac{4}{3}\pi^2 + \frac{9}{2} \right]
  + \frac{1}{\eps} \, \biggl[
    -\frac{4}{3} \lqm^3
  \nonumber \\* & \qquad
    + 4 \lqm^2
    + \left(\frac{2}{3}\pi^2-17\right) \lqm
    + 12\zeta_3 - \frac{7}{3}\pi^2 + \frac{85}{4} \biggr]
  \nonumber \\ & \quad
  + \frac{2}{3} \lqm^4
  - \frac{8}{3} \lqm^3
  + \left(\frac{\pi^2}{3} + 17\right) \lqm^2
  \nonumber \\ & \quad
  + \left(-36\zeta_3 + \frac{2}{3}\pi^2 - \frac{101}{2}\right) \lqm
  + \frac{107}{90}\pi^4
  + \frac{184}{3}\zeta_3
  \nonumber \\* & \quad
  - \frac{59}{12}\pi^2
  + \frac{599}{8}
  \Biggr\}
  + \Oc(\eps) + \Oc\!\left(\frac{M^2}{Q^2}\right)
  ,
\end{align}
where $\alpha$ and $\alpha'$ are the couplings of the $SU(2)$ and
$U(1)$ gauge groups, respectively, and $\lqm = \ln(Q^2/M^2)$.
The $1/\eps^3$ pole is due to the infrared divergence.
The same diagram with line~6 massless yields the contribution
\begin{align}
\label{eq:LAm5res}
  F_{\vr,\LA}^{M_6=0} &=
    C_F
    \, \frac{\alpha \alpha'}{(4\pi)^2}
    \left(\frac{\mu^2}{M^2}\right)^{2\eps} S_\eps^2
    \, \Biggl\{
  \frac{1}{2\eps^2}
  \nonumber \\* & \qquad
  + \frac{1}{\eps} \left[
    - \lqm^2 + 3 \lqm - \frac{2}{3}\pi^2 - \frac{11}{4} \right]
  + \frac{1}{6} \lqm^4
  \nonumber \\ & \qquad
  + \left(\frac{2}{3}\pi^2 - 1\right) \lqm^2
  + \left(-24\zeta_3 - \pi^2 + \frac{11}{2}\right) \lqm
  \nonumber \\ & \qquad
  + \frac{13}{45}\pi^4 + 46\zeta_3 + \frac{13}{12}\pi^2 - \frac{41}{8}
  \Biggr\}
  \nonumber \\* & \quad
  + \Oc(\eps) + \Oc\!\left(\frac{M^2}{Q^2}\right)
  .
\end{align}
Note that only the linear logarithm and the non-loga\-rith\-mic constant
at order~$\eps^0$ of this result differ from the case~(\ref{eq:LAres})
with two massive gauge bosons (and, of course, the different prefactor
$C_F \, \alpha \alpha'$ instead of $C_F^2 \, \alpha^2$).

When either of the two gauge bosons in the \emph{non-planar vertex
diagram} of figure~\ref{fig:feynAb}b) is massless
(cf. figure~\ref{fig:scalarNP} for the line numbering), the
contribution is
\begin{align}
\label{eq:NPm5res}
  F_{\vr,\NP}^{M_5=0} &= F_{\vr,\NP}^{M_6=0} =
    C_F
    \, \frac{\alpha \alpha'}{(4\pi)^2}
    \left(\frac{\mu^2}{M^2}\right)^{2\eps} S_\eps^2
    \, \Biggl\{
  \nonumber \\* & \qquad
  \frac{1}{\eps} \left[
    -\frac{2}{3} \lqm^3 + 4 \lqm^2 - 12 \lqm
    - 12\zeta_3 + \pi^2 + 14 \right]
  \nonumber \\* & \qquad
  + \frac{1}{2} \lqm^4
  - 4 \lqm^3
  + \left(-\frac{5}{3}\pi^2 + 22\right) \lqm^2
  \nonumber \\* & \qquad
  + \left(56\zeta_3 + \frac{11}{3}\pi^2 - 68\right) \lqm
  - \frac{67}{90}\pi^4 - 90\zeta_3
  \nonumber \\* & \qquad
  - 4\pi^2 + 96
  \Biggr\}
  + \Oc(\eps) + \Oc\!\left(\frac{M^2}{Q^2}\right)
  \,.
\end{align}
The \emph{Mercedes--Benz graph} in figure~\ref{fig:feynAb}c) gives the
contribution
\begin{align}
\label{eq:BEm4res}
  F_{\vr,\BE}^{M_3=0} &=
    C_F
    \, \frac{\alpha \alpha'}{(4\pi)^2}
    \left(\frac{\mu^2}{M^2}\right)^{2\eps} S_\eps^2
    \, \Biggl\{
  \frac{1}{2\eps^2}
  \nonumber \\* & \qquad
  + \frac{1}{\eps} \, \biggl[
    - \lqm^2
    + \left(-\frac{2}{3}\pi^2 + 7\right) \lqm
    + 4\zeta_3 + \frac{\pi^2}{3}
  \nonumber \\* & \qquad\qquad
    - \frac{53}{4} \biggr]
  + \lqm^3
  + \left(\frac{2}{3}\pi^2 - 9\right) \lqm^2
  \nonumber \\ & \qquad
  + \left(-4\zeta_3 - 3\pi^2 + \frac{89}{2}\right) \lqm
  - \frac{13}{90}\pi^4 + 16\zeta_3
  \nonumber \\* & \qquad
  + \frac{79}{12}\pi^2 - \frac{655}{8}
  \Biggr\}
  + \Oc(\eps) + \Oc\!\left(\frac{M^2}{Q^2}\right)
  ,
\end{align}
when line~3 (cf. figure~\ref{fig:scalarBE}) is massless, and
\begin{align}
\label{eq:BEm3res}
  F_{\vr,\BE}^{M_4=0} &=
    C_F
    \, \frac{\alpha \alpha'}{(4\pi)^2}
    \left(\frac{\mu^2}{M^2}\right)^{2\eps} S_\eps^2
    \, \Biggl\{
  -\frac{2}{\eps^3}
  + \frac{1}{\eps^2} \left[
    2 \lqm - \frac{5}{2} \right]
  \nonumber \\* & \qquad
  + \frac{1}{\eps} \left[
    - 2 \lqm^2 + 7 \lqm - \frac{\pi^2}{3} - \frac{53}{4} \right]
  + \frac{4}{3} \lqm^3
  \nonumber \\ & \qquad
  + \left(\frac{\pi^2}{3} - 9\right) \lqm^2
  + \left(8\zeta_3 - \frac{7}{3}\pi^2 + \frac{73}{2}\right) \lqm
  \nonumber \\ & \qquad
  + \frac{11}{45}\pi^4 - \frac{32}{3}\zeta_3 + \frac{17}{4}\pi^2
  - \frac{479}{8}
  \Biggr\}
  \nonumber \\* & \quad
  + \Oc(\eps) + \Oc\!\left(\frac{M^2}{Q^2}\right)
  ,
\end{align}
when line~4 is massless.

The \emph{vertex correction with fermion self-energy} of the diagram
in figure~\ref{fig:feynAb}d) yields
\begin{align}
\label{eq:fcm5res}
  F_{\vr,\fc}^{M_3=0} &=
    C_F
    \, \frac{\alpha \alpha'}{(4\pi)^2}
    \left(\frac{\mu^2}{M^2}\right)^{2\eps} S_\eps^2
    \, \Biggl\{
  \frac{2}{\eps^3}
  + \frac{1}{\eps^2} \left[
    -2 \lqm + \frac{5}{2} \right]
  \nonumber \\* & \qquad
  + \frac{1}{\eps} \left[
    2 \lqm^2 - 7 \lqm + \frac{\pi^2}{3} + \frac{53}{4} \right]
  - \frac{4}{3} \lqm^3
  + 7 \lqm^2
  \nonumber \\ & \qquad
  + \left(\frac{\pi^2}{3} - \frac{53}{2}\right) \lqm
  - \frac{40}{3}\zeta_3 - \frac{13}{12}\pi^2 + \frac{355}{8}
  \Biggr\}
  \nonumber \\* & \quad
  + \Oc(\eps) + \Oc\!\left(\frac{M^2}{Q^2}\right)
  ,
\end{align}
with line~3 (cf. figure~\ref{fig:scalarfc}) massless and
\begin{align}
\label{eq:fcm3res}
  F_{\vr,\fc}^{M_5=0} &=
    C_F
    \, \frac{\alpha \alpha'}{(4\pi)^2}
    \left(\frac{\mu^2}{M^2}\right)^{2\eps} S_\eps^2
    \, \Biggl\{
  -\frac{1}{2\eps^2}
  \nonumber \\* & \qquad
  + \frac{1}{\eps} \left[
    \lqm^2 - 3 \lqm + \frac{2}{3}\pi^2 + \frac{13}{4} \right]
  - \lqm^3
  + 5 \lqm^2
  \nonumber \\* & \qquad
  - \frac{33}{2} \lqm
  - 4\zeta_3 + \frac{\pi^2}{12} + \frac{163}{8}
  \Biggr\}
  \nonumber \\* & \quad
  + \Oc(\eps) + \Oc\!\left(\frac{M^2}{Q^2}\right)
  ,
\end{align}
with line~5 massless. The only difference of~(\ref{eq:fcm3res}) with
respect to the purely massive result~(\ref{eq:fcres}) is in the
non-loga\-rith\-mic constant at order~$\eps^0$.

The self-energy diagram in figure~\ref{fig:feynSigmaAb}a) contributes
\begin{align}
\label{eq:T1m2res}
  \Sigma_\Tone^{M_2=0} = \Sigma_\Tone^{M_3=0} &=
    C_F
    \, \frac{\alpha \alpha'}{(4\pi)^2}
    \left(\frac{\mu^2}{M^2}\right)^{2\eps} S_\eps^2
    \left(
  \frac{1}{2\eps}
  - \frac{3}{4}
  \right)
  \nonumber \\* & \qquad
  + \Oc(\eps)
  \,,
\end{align}
when either of its two gauge bosons is massless.
The two contributions of the self-energy diagram in
figure~\ref{fig:feynSigmaAb}b) are
\begin{align}
\label{eq:T2m4res}
  \Sigma_\Ttwo^{M_2=0} &=
    C_F
    \, \frac{\alpha \alpha'}{(4\pi)^2}
    \left(\frac{\mu^2}{M^2}\right)^{2\eps} S_\eps^2
  \left(
  -\frac{1}{2\eps^2}
  + \frac{3}{4\eps}
  - \frac{\pi^2}{4} - \frac{1}{8}
  \right)
  \nonumber \\* & \qquad
  + \Oc(\eps)
  \,,
\end{align}
with the gauge boson in the outer loop massless, and
\begin{align}
\label{eq:T2m2res}
  \Sigma_\Ttwo^{M_4=0} &=
    C_F
    \, \frac{\alpha \alpha'}{(4\pi)^2}
    \left(\frac{\mu^2}{M^2}\right)^{2\eps} S_\eps^2
    \left(
  \frac{1}{2\eps^2}
  - \frac{1}{4\eps}
  + \frac{\pi^2}{4} - \frac{1}{8}
  \right)
  \nonumber \\* & \qquad
  + \Oc(\eps)
  \,,
\end{align}
with the gauge boson in the inner loop massless. Note that
(\ref{eq:T2m2res}) differs from the purely massive
result~(\ref{eq:T2res}) only at order~$\eps^0$.

The self-energy diagram in figure~\ref{fig:feynSigmaAb}c) has no
corresponding contribution with one massive and one massless gauge
boson, because these integrals vanish in dimensional regularization.

According to~(\ref{eq:F2cont}), the product of the massless one-loop
vertex correction $F_{\vr,1}^{M=0}$ and the (massive) one-loop
self-energy correction $\Sigma_1$~(\ref{eq:Sigma1res}) is needed as
well. The missing piece is well known:
\begin{multline}
\label{eq:Fv1hres}
  F_{\vr,1}^{M=0} =
  \frac{\alpha'}{4\pi} \left(\frac{\mu^2}{Q^2}\right)^\eps S_\eps
  \, \Biggl\{
    -\frac{2}{\eps^2} - \frac{3}{\eps} + \frac{\pi^2}{6} - 8
  \\*
    + \eps \left( \frac{14}{3}\zeta_3 + \frac{\pi^2}{4} - 16 \right)
  \Biggr\}
  + \Oc(\eps^2)
  \,.
\end{multline}
The prefactor has to be expanded as
\[
  \left(\frac{\mu^2}{Q^2}\right)^\eps =
  \left(\frac{\mu^2}{M^2}\right)^\eps \left(
    1 - \eps \lqm + \frac{\eps^2}{2} \lqm^2 - \frac{\eps^3}{6} \lqm^3
  \right)
  + \Oc(\eps^4)
\]
in order to match the prefactor of the other contributions.

The contributions to the $SU(2) \times U(1)$ form factor with one
massive and one massless gauge boson may now be added together
in analogy with~(\ref{eq:F2CFres}):
\begin{align}
\label{eq:F2intres}
  F_{2,\alpha\alpha'} &=
    F_{\vr,\LA}^{M_5=0} + F_{\vr,\LA}^{M_6=0}
    + F_{\vr,\NP}^{M_5=0} + F_{\vr,\NP}^{M_6=0}
  \nonumber \\* & \qquad
    + 2\,F_{\vr,\BE}^{M_3=0} + 2\,F_{\vr,\BE}^{M_4=0}
    + 2\,F_{\vr,\fc}^{M_3=0} + 2\,F_{\vr,\fc}^{M_5=0}
  \nonumber \\* & \qquad
    + \Sigma_\Tone^{M_2=0} + \Sigma_\Tone^{M_3=0}
    + \Sigma_\Ttwo^{M_2=0} + \Sigma_\Ttwo^{M_4=0}
  \nonumber \\* & \qquad
    + F_{\vr,1}^{M=0} \Sigma_1
  \nonumber \\ &=
    C_F
    \, \frac{\alpha \alpha'}{(4\pi)^2}
    \left(\frac{\mu^2}{M^2}\right)^{2\eps} S_\eps^2
    \, \Biggl\{
  \nonumber \\* & \qquad
  \frac{1}{\eps^2} \left[
    2 \lqm^2 - 6 \lqm + \frac{4}{3}\pi^2 + 7 \right]
  + \frac{1}{\eps} \, \biggl[
    -\frac{8}{3} \lqm^3 + 12 \lqm^2
  \nonumber \\* & \qquad\qquad
    + \left(-\frac{2}{3}\pi^2 - 32\right) \lqm
    - 4\zeta_3 + \pi^2 + 34 \biggr]
  \nonumber \\ & \qquad
  + \frac{11}{6} \lqm^4
  - 11 \lqm^3
  + \left(-\frac{\pi^2}{3} + 49\right) \lqm^2
  \nonumber \\ & \qquad
  + \Bigl(60\zeta_3 - 3\pi^2 - 111\Bigr) \, \lqm
  + \frac{17}{90}\pi^4 - 102\zeta_3
  \nonumber \\* & \qquad
  + \frac{47}{6}\pi^2 + 117
  \Biggr\}
  + \Oc(\eps) + \Oc\!\left(\frac{M^2}{Q^2}\right)
  .
\end{align}
The infrared-convergent two-loop interference term of equation~(6)
in~\cite{Feucht:2004rp} results from~(\ref{eq:F2intres}) after
subtraction of the massive times the massless one-loop form factor:
$F_{2,\alpha\alpha'} - (F_{\vr,1} + \Sigma_1) \cdot F_{\vr,1}^{M=0}$.

\bibliographystyle{epjc}
\bibliography{references}

\end{document}